\documentclass[11pt,oneside,letterpaper]{article}
\usepackage{inputenc}
\usepackage{amssymb}
\usepackage{amsmath,bm}
\usepackage[dvips]{graphicx}
\usepackage{setspace}
\usepackage{fancyhdr}
\usepackage[dvipsnames]{xcolor}
\usepackage{ifpdf}
\usepackage{graphicx}
\usepackage{rotating}
\usepackage{comment}
\usepackage[margin=2.5cm]{geometry}
\usepackage[colorlinks=true, linkcolor  = blue,urlcolor=blue,citecolor=violet]{hyperref}
\usepackage{soul}
\usepackage{relsize}
\usepackage{tikz}
\usepackage{float}
\usepackage{empheq}
\usepackage{bbold}
\usepackage{cite}
\usepackage{etoolbox}
\usepackage{extarrows}
\def\sub{\scriptscriptstyle}
\def\e{{\epsilon}}

\def\F{\mathcal{F}}
\def\Ups{\Upsilon}
\def\R{\mathcal R}
\def\a{\alpha}
\def\b{\beta}
\def\d{\delta}

\def\0{\scriptscriptstyle{0}}
\newcommand{\bi}{\begin{itemize}}
\newcommand{\ei}{\end{itemize}}
\newcommand{\bea}{\begin{eqnarray}}
\newcommand{\eea}{\end{eqnarray}}
\newcommand{\be}{\begin{equation}}
\newcommand{\ee}{\end{equation}}

\newcommand{\nn}{\nonumber}
\newcommand{\dd}{\mathrm{d}}

\numberwithin{equation}{section}
\allowdisplaybreaks  

\begin{document}
\def\biblio{}



\onehalfspacing

\begin{center}

~
\vskip4mm
{{\huge {
\quad 
Boundary timelike Liouville theory: bulk 1-point \& boundary 2-point functions
 }
  }}
\vskip5mm

\vskip2mm

\vskip10mm

Teresa Bautista$^1$  ~and ~Aditya Bawane$^2$ \\ 

\end{center}
\vskip4mm
\begin{center}
{
\footnotesize
{$^1$Department of Mathematics, King's College London, Strand, London WC2R 2LS, UK \\
$^2$Faculty of Physics, University of Warsaw,
ul. Pasteura 5, 02-093 Warsaw, Poland\\
}}
\end{center}
\begin{center}
{\textsf{\footnotesize{
teresa.bautista@kcl.ac.uk, aditya.bawane@fuw.edu.pl
}} } 
\end{center}
\vskip5mm

\vspace{4mm}
 
\vspace*{0.6cm}

\vspace*{1.5cm}
\begin{abstract}
\noindent
We consider timelike Liouville theory with FZZT-like boundary conditions. The bulk one-point and boundary two-point structure constants on a disk are derived using bootstrap. We find that these structure constants are not the analytic continuations of their spacelike counterparts. 
\end{abstract}

\newpage
\setcounter{page}{1}
\pagenumbering{arabic}

\tableofcontents

\section{Introduction}

Timelike Liouville gravity is a two-dimensional model of gravity built upon Liouville conformal field theory in its non-unitary or timelike regime. Starting from the Euclidean path integral of gravity in two dimensions with a cosmological constant and coupled to unitary $c_m\geq 25$ conformal matter, in the conformal gauge, the effective action for the conformal factor of the metric is the timelike Liouville action \cite{Polyakov:1981:QuantumGeometryBosonic}, and the conformal factor takes  the role of the timelike Liouville field. In such a setting, gravity is a conformal field theory.

The word timelike \cite{Strominger:2003fn} is due to the fact that the kinetic term of the Liouville field in the action appears with an additional minus sign, so that the Liouville direction is a timelike direction in a Lorentzian-signature field space \cite{Das:1988ds}. This implies that the theory is non-unitary; accordingly its central charge is $c_L\leq 1$. Despite requiring a more difficult quantisation, this feature makes this theory a very interesting toy model of higher-dimensional gravity: it reproduces the well-known wrong-sign kinetic term problem of the Weyl factor of the metric in Einstein-Hilbert gravity, identified  already more  than forty years ago  \cite{Gibbons:1978ac}, which entails an action unbounded from below and hence an ill-defined Euclidean path integral of gravity. Timelike Liouville theory is therefore a very suited model to address this issue, since it allows to tackle it with all the bootstrap techniques of conformal field theories.

Another advantage of this theory is that it can be coupled to unitary conformal matter. Diffeomorphism invariance results in conformal symmetry in the conformally-flat gauge. As a result, the total central charge of the theory has to vanish, i.e. $c_L+c_m=26$. Since $c_L\leq 1$, the matter sector must have $c_m\geq 25$. This is in contrast to what happens in the so-called \textit{spacelike} Liouville gravity. In the spacelike regime, the Liouville field has the right-sign kinetic term, hence the theory is unitary and is modeled by the standard Liouville CFT with $c_L\geq 25$. As a CFT, spacelike Liouville theory was already solved many years ago, and many of its properties are long well understood (for reviews see \cite{Nakayama:2004vk,Ribault:2014hia}). As a theory of gravity with a cosmological constant, it was very much explored thanks to its connections (see \cite{Knizhnik:1988ak,Moore:1991ir,Moore:1991ag,Belavin:2014hsa,Belavin:2008kv} among many others) to discrete models of two-dimensional gravity. However, as a lower-dimensional theory of Einstein-Hilbert gravity in the conformal gauge, it needs to be coupled to a (possibly) non-unitary $c_m\leq 1$ matter (so that $c_L+c_m=26$), and hence makes for a more exotic model than its timelike counterpart.

As a CFT, progress in solving timelike Liouville theory was achieved during the past decade. In particular, a 3-point structure constant which solves the degenerate bootstrap equations \cite{Teschner:1995yf} was computed \cite{Zamolodchikov:2005fy,Kostov:2005kk,Kostov:2007:NonRational2DQuantum-1,Kostov:2006zp,Harlow:2011ny}, and later proven to satisfy all crossing-symmetry constraints \cite{Ribault:2015sxa}. These results were consequently used to show that timelike Liouville CFT can accommodate a unitary theory of gravity coupled to conformal matter, by identifying the allowed gravitational spectrum and by further showing its consistency with the conformal symmetry constraints \cite{Bautista:2019jau,Bautista:2020obj}. 
Given the consistency and viability of this theory, both as a CFT and as a theory of gravity, it is now time to explore its generalisations. One such generalisation consists of placing the theory on a space with boundaries.

 Boundary spacelike Liouville theory was thoroughly studied in the past and proven to be very fruitful. As a boundary CFT (BCFT), all of its data has been computed:  two bulk 1-point function solutions were found, the FZZT \cite{Fateev:2000ik,Teschner:2000md} and the ZZ branes \cite{Zamolodchikov:2001ah} corresponding to Neumann and Dirichlet boundary conditions respectively, the boundary 2-point function was found in \cite{Fateev:2000ik,Teschner:2000md}, the boundary 3-point function was determined in \cite{Ponsot:2001ng}, and the bulk-boundary 2-point function in \cite{Hosomichi:2001xc}.

The two 1-point function solutions, the FZZT and ZZ branes, have played an important role in  several developments in lower-dimensional string theory on time-dependent backgrounds \cite{Sen:2002nu,Sen:2004zm}. 
Furthermore, ZZ branes have been relevant to understanding the possible discrete nature of 2d gravity through their connection to matrix models \cite{McGreevy:2003kb,McGreevy:2003ep,Douglas:2003up}.
Further applications of the spacelike Liouville BCFT conformal data to two-dimensional  quantum gravity can be found in \cite{Kostov:2002uq,Kostov:2003uh} (reviews include \cite{Nakayama:2004vk,Kostov:2003uh,Douglas:2003up}). The success of these developments in the spacelike regime motivates the study of boundary conditions in timelike Liouville theory. 

In this paper, we  study boundary timelike Liouville theory. Concretely, we compute the bulk 1-point structure constant analogous to the spacelike FZZT solution, given in \eqref{1-pnt-fn-normalised}, and the boundary 2-point structure constant, given in \eqref{bdry2pt_TL}.  The main outcome of our work is that these two structure constants do not correspond to the analytic continuations of their spacelike counterparts. We employ familiar bootstrap techniques \cite{Teschner:1995yf,Fateev:2000ik,Zamolodchikov:2005fy}, such as using degenerate operators to derive shift equations. These have been successfully  used in the spacelike regime and for the sphere 3-point structure constant in the timelike theory.

In the past there have been some attempts to solve for such CFT data in timelike CFTs \cite{Gutperle:2003xf,Fredenhagen:2004cj,Gaberdiel:2004na}. Most of these works however work in some approximation or some particular case, such as limiting to the $c_L=1$ Liouville theory or setting the cosmological constant to zero. To the best of our knowledge, boundary timelike Liouville theory at generic central charge or cosmological constant has not been explored in the past.

The outline of the paper is the following. In section \ref{section:TL} we give an introduction to timelike Liouville theory, with its basic correlators in the full complex plane. In section \ref{sec:timelike boundary Liouville} we then present our results for boundary timelike Liouville on the disk or the upper half plane. Concretely, in section \ref{sec:1-pnt-fn} we present our result for the bulk 1-point function, and in section \ref{sec:bdry-2pnt-fn} we present our result for the boundary 2-point function. We conclude in section \ref{sec:discussion} with a discussion of our results, a comparison with those in spacelike Liouville, and some outlook for the future. Finally, several appendices cover some of the more detailed computations required in the introduction and bulk of the work.

\section{Timelike Liouville theory}\label{section:TL}

In this first section, we give a brief introduction to the main elements of timelike Liouville theory on the sphere or the plane, thereby setting up our notation. Reviews can be found in \cite{Harlow:2011ny,Ribault:2014hia,Bautista:2019jau}. 

Timelike Liouville theory consists of an interacting timelike scalar $\chi$, the Liouville field, with action
\begin{equation}\label{action-timelike-Liou}
S_{tL}[\chi]=\frac{1}{4\pi}\,\int d^2z\sqrt{h}\,\left( - (\nabla\chi)^2-q \,R_h\,\chi+4\pi\mu\,e^{2\beta\chi}\right),
\end{equation}
where $\mu$ is the cosmological constant, $q$ is the so-called background charge and $\beta$ is the Liouville coupling constant. Besides the exponential interaction, it exhibits a linear coupling of the field to the fixed background curvature $R_h$ weighted by the background charge. Despite the cosmological constant being dimensionful, this action exhibits Weyl invariance, where the Weyl transformation shifts the Liouville field linearly:
\be
h\rightarrow \,e^{2\sigma(z)}\,h,\qquad \chi\rightarrow \chi-q\sigma(z).
\ee

When the fiducial metric $h$ is the flat metric $ds^2=dz\,d\bar z$, the action becomes\footnote{So written, the action diverges upon evaluating it on its solutions. To regularise it, we can place it on a disk and introduce the corresponding boundary terms, so that the large radius limit is finite \cite{Zamolodchikov:1995aa}. We disregard these terms here since they are irrelevant for the presentation.}
\begin{align}
S_{tL}[\chi]
&=\frac{1}{2\pi}\,\int dz\,d\bar z\,\left( - \partial\chi\bar\partial\chi+\pi\mu\,e^{2\beta\chi}\right).
\end{align}
The Weyl symmetry then descends to conformal symmetry and Liouville theory becomes a CFT. At the quantum level it has been shown to be a solution to the bootstrap equations and constitutes a consistent CFT  on all orientable Riemann surfaces \cite{Ribault:2015sxa}. Its central charge is parametrised by the background charge as
\be
c=1-6q^2.
\ee
We will focus on real actions and hence on $q\in\mathbb R$, so the central charge is mostly negative $c\leq 1$. 

Given the linear transformation of the field, the natural primaries of this conformal field theory are vertex operators 
\be
V_\alpha=e^{-2\alpha\chi},
\ee
where $\alpha$ is called the Liouville charge. Remarkably, vertex operators in such an interacting theory have the same anomalous dimension as in free theory, so that
\be
\Delta_\a=\bar\Delta_\a=\alpha(\alpha-q),
\ee
the $-\alpha q$ contribution being classical and the $\alpha^2$ contribution being anomalous. In particular, the cosmological constant operator in the action $V_{-\beta}=e^{2\beta\chi}$ has dimension $\Delta_\b=\bar\Delta_\b=\beta(\beta+q)$. Given that the action has to be conformally-invariant, this operator must have dimensions $(1,1)$, which then implies the well-known relation between the coupling constant and the background charge
\be
q=\frac{1}{\beta}-\beta.
\ee
The semiclassical limit corresponds to $q\rightarrow\infty$ or $\beta\rightarrow 0$.

The Liouville charge $\alpha$ parametrises the spectrum of the theory. While it is a priori complex, its range is constrained by conformal invariance and by any further physical requirements the theory may need to satisfy. Crossing symmetry of 4-point functions constrains intermediate or \textit{internal} states to have $\alpha\in\mathbb R$ \cite{Ribault:2015sxa}, which gives a spectrum of internal conformal dimensions bounded from below with minimum at $\Delta_{q/2}=\bar\Delta_{q/2}=-q^2/4$. However, the charges of the insertions of the 4-point function or \textit{external} states can be analytically continued outside of this range while keeping the 4-point function crossing symmetric, so that the actual spectrum of  external charges is not constrained by crossing. When the Liouville field corresponds to the conformal factor of the metric, i.e. when timelike Liouville is a theory of gravity, diffeomorphism invariance needs to be further imposed. This then restricts the range of $\alpha$, but in a way that is compatible with unitarity of the whole gravity+matter  theory. For more details see \cite{Bautista:2019jau,Bautista:2020obj}. 

The timelike Liouville action \eqref{action-timelike-Liou} is related to the well-known spacelike Liouville action,
\be
S_{sL}[\phi]=\frac{1}{4\pi}\int d^2z\sqrt{h}\left((\nabla\phi)^2+Q\,R_h\,\phi+4\pi\mu\,e^{2b\phi}\right),
\ee
by the analytic continuation
\be\label{analytic-continuation}
\phi=i\chi,\qquad Q=i q,\qquad b=-i\beta,\qquad a=i\alpha.
\ee
This analytic continuation ensures both actions can be real. Spacelike Liouville theory has central charge $c=1+6Q^2$ with $c\geq 1$, and its primaries are given by $V_a=e^{2a\phi}$ with $\Delta_a=\bar\Delta_a=a(Q-a)$. The spectrum is unitary, consisting of Liouville charges $a=\frac{Q}{2}+i P$ with Liouville momentum $P\in\mathbb R$. Spacelike Liouville theory is a well-established CFT, its bulk correlators as well as its boundary state solutions are well known. Unfortunately, the analytic continuation \eqref{analytic-continuation} of some of this data to the timelike regime is not well-defined. The corresponding timelike CFT data then needs to be found independently. We review this next for the bulk correlators.

\subsection{Correlators}\label{correlators}

As in any CFT, higher-point correlators can be determined from the 2- and 3-point functions of the theory. 
 The timelike Liouville 3-point function on the complex plane reads
 \begin{align}\label{TL-3pnt-fn}
 \langle V_{\alpha_1}(z_1)\,V_{\alpha_2}(z_2)\,V_{\alpha_3}(z_3)\rangle=\frac{   C(\alpha_1,\alpha_2,\alpha_3)}{|z_{12}|^{2(\Delta_t-2\Delta_3)}\,|z_{13}|^{2(\Delta_t-2\Delta_2)}\,|z_{23}|^{2(\Delta_t-2\Delta_1)}},
 \end{align}
where we use the notation $\Delta_i$ to indicate $\Delta_{\a_i}$, $\Delta_t=\sum \Delta_i$ is the sum of all dimensions, and the structure constant is \cite{Zamolodchikov:2005fy,Kostov:2005kk,Kostov:2007:NonRational2DQuantum-1,Kostov:2006zp}
\begin{align}\label{TL-structure-cntt}
   C(\alpha_1,\alpha_2,\alpha_3)=-\frac{1}{2\beta}\left(\pi\mu\,\gamma(-\beta^2)\,\beta^{2+2\beta^2}\right)^{\frac{\alpha_{\scriptscriptstyle{t}}-q}{\beta}}\,
\frac{\Upsilon_\beta(\beta-q+\alpha_{\scriptscriptstyle{t}})}{\Upsilon_\beta(\beta)}
\prod\limits_{i=1}^3
\frac{\Upsilon_\beta(\alpha_{\scriptscriptstyle{t}}-2\alpha_i+\beta)}{\Upsilon_\beta(\beta+2\alpha_i)},
\end{align}
with $\alpha_t=\sum\alpha_i$  and $\gamma(x):=\Gamma(x)/\Gamma(1-x)$. The Upsilon function $\Upsilon_\beta(x)$ \cite{Zamolodchikov:1995aa} has  a simple integral definition for $\text{Re}( x) \in (0,\text{Re} (\b^{-1}+\b))$:
\begin{equation}
\ln \Upsilon_\b(x)
= \int_0^{\infty} \frac{\dd t}{t} \left[
\left( \frac{\b^{-1}+\b}{2} - x \right)^2 e^{- 2t}
- \frac{\sinh^2 \left( \left( \frac{\b^{-1}+\b}{2} - x \right) t\right)}{\sinh (\b t)\, \sinh\left( \frac{t}{\b}\right)}
\right].
\end{equation} 
This formula admits an analytic continuation to $x \in \mathbb C$, and can also be represented by an infinite product:
\begin{equation}
\Upsilon_\b(x)
= \lambda^{\left( \frac{1}{2}(\b^{-1}+\b) - x \right)^2}
\prod_{m,n \in \mathbb N} f\left( \frac{\frac{\b^{-1}+\b}{2} - x}{\frac{\b^{-1}+\b}{2} + m \b + n \b^{-1}} \right),
\qquad
f(x) = (1 - x^2) \, \e^{x^2},
\end{equation} 
where $\lambda$ is some constant.
Importantly, this function satisfies shift relations with shift parameters $\beta,\beta^{-1}$; see appendix \ref{app:special functions} for more properties of this function. 

As mentioned above, this structure constant does not follow from the analytic continuation \eqref{analytic-continuation} of the well-known structure constant of spacelike Liouville given by the DOZZ formula \cite{Dorn:1994xn,Zamolodchikov:1995aa}, since such a continuation diverges \cite{Zamolodchikov:2005fy}. Instead, this structure constant was found as an independent solution to the conformal bootstrap constraints: the equations that follow from the associativity property of the OPE, or equivalently from crossing symmetry, and which must be satisfied for any CFT. 

Concretely, the strategy consists of looking for a solution of a subset of the bootstrap constraints  \cite{Teschner:1995yf}, sometimes called the degenerate equations. The degenerate equations are shift equations  for the Liouville structure constants that follow from crossing symmetry of the 4-point function where one of the four insertions is a level-2 degenerate field. Degenerate fields $V_{\langle m,n\rangle}$, at level $m\, n$, are parametrised by two positive integers $(m,n)$, and have charges $\a_{m,n}=\frac{1-m}{2\beta}-\frac{(1-n)\b}{2}$. They are the primaries of degenerate representations, i.e., quotients of Verma modules. For a pedagogical reference see \cite{Ribault:2014hia}.

 Two shift equations are then obtained for each of the two degenerates at level 2: $V_{\langle1,2\rangle}$ and $V_{\langle2,1\rangle}$.  These are only two equations and are effectively linear in the 3-point structure constant, so they are much easier to solve than the infinite set of general bootstrap constraints which are quadratic in the structure constant and are integral equations. The solution found for the timelike regime \cite{Zamolodchikov:2005fy,Kostov:2005kk,Kostov:2007:NonRational2DQuantum-1,Kostov:2006zp}  was later proven to solve all of the bootstrap equations numerically \cite{Ribault:2015sxa}, thus confirming it is the correct  timelike 3-point structure constant. 

The shift equations do not determine the normalization of the 3-point structure constant.
	Our choice of normalization is based on the so-called Coulomb gas method or perturbative method, which we explain in appendix \ref{app:DF-integrals}. In appendix \ref{app-bulk-correls}  we review in detail the derivation of the bulk timelike structure constants, with particular emphasis on separating the  normalization-independent factors, from those that follow from fixing the normalization. 

One last comment about the normalization. In the semiclassical limit $\b\rightarrow 0$, $\gamma(-\beta^2) \rightarrow -1/\beta^2$,  so the argument of the normalization parenthesis  in \eqref{TL-structure-cntt}, 
\be
\left(\pi\mu\,\gamma(-\beta^2)\,\beta^{2+2\beta^2}\right)^{\frac{\alpha_{\scriptscriptstyle{t}}-q}{\beta}},
\ee
is negative for $\mu>0$. 
In that case, a phase factor  $\exp\{-i\pi\frac{\alpha_{\scriptscriptstyle{t}}-q}{\beta}\}$ should be included \cite{Harlow:2011ny}.

\vspace{5mm}

The general form of the 2-point function is
\be\label{2pntfn-gen}
\langle V_{\alpha_1}(z_1)\,V_{\alpha_2}(z_2)\rangle=2\pi\,\frac{   G(\alpha_1)\,\left[\delta(\alpha_1-\alpha_2)+   \mathcal R(\alpha_2)\delta(q-\alpha_1-\alpha_2)\right]}{|z_{12}|^{2(\Delta_1+\Delta_2)}}.
\ee
$G(\a)$ is the 2-point function structure constant, and the coefficient $\R(\a)$ is the so-called reflection coefficient. This reflection can be understood as coming from the invariance of conformal dimensions $\Delta_\a=\bar\Delta_\a=\alpha(\alpha-q)$ under $\alpha\rightarrow q-\alpha$. This implies that the pair of operators $V_\alpha$ and $V_{q-\alpha}$ have the same dimension, and  must therefore be related by a reflection coefficient $   R(\alpha)$, such that
\be\label{reflection-rel-ops}
V_\alpha=   \mathcal R(\alpha)\,V_{q-\alpha},
\ee
and which satisfies $   \R(\alpha)\,   \R(q-\alpha)=1$.

Given \eqref{reflection-rel-ops}, the reflection coefficient can be obtained from the 3-point structure constant \eqref{TL-structure-cntt} by reflecting one of the operators, and is given by
\be\label{reflection-coef-timelike}
\R(\alpha)=\left(\pi\mu\,\gamma(-\beta^2)\right)^{\frac{2\alpha-q}{\beta}}\,
\frac{\Gamma\left(\beta(2\alpha-q)\right)\,\Gamma\left(\beta^{-1}(q-2\alpha)\right)}{\Gamma\left(\beta(q-2\alpha)\right)\,\Gamma\left(\beta^{-1}(2\alpha-q)\right)}.
\ee
Notice that this expression is independent of the choice of $\a_i$-independent normalization chosen for the 3-point structure constant.

As opposed to the 3-point function, the 2-point function structure constant and reflection coefficient do have a good analytic continuation \eqref{analytic-continuation} between the spacelike and timelike regimes, and \eqref{reflection-coef-timelike} coincides with the analytic continuation of the spacelike reflection coefficient, up to a minus sign.  Given that in the spacelike regime the 2-point structure constant and the reflection coefficient are taken to coincide, it is natural to take this convention in the timelike regime as well, $G(\a)=\R(\a)$, so that the 2-point function becomes
\be\label{2pnt-fn-timelike}
\langle V_{\alpha_1}(z_1)\,V_{\alpha_2}(z_2)\rangle=2\pi\,\frac{  \R(\alpha_1)\,\delta(\alpha_1-\alpha_2)+\delta(q-\alpha_1-\alpha_2)}{|z_{12}|^{2(\Delta_1+\Delta_2)}}.
\ee
This 2-point function then coincides with the analytic continuation of the spacelike one up to an overall minus sign.
The overall factor of $2\pi$ arises in the spacelike 2-point function by defining the latter using the limit\footnote{This limit can be verified by using the expression for the spacelike structure constant, the DOZZ formula, and using the limit $\lim\limits_{\epsilon\rightarrow 0} \frac{\epsilon}{\epsilon^2-x^2}=\pi \,\delta(x)$ and the asymptotic behaviour of the Upsilon function $\lim\limits_{x\rightarrow 0}\Upsilon_b(x)=\Upsilon_b(b)\,x$.\label{footnote-limit-2-to-3pnt-fn}}
\be\label{2pnt-fn-from-3pnt-fn-space}
\langle V_{a_1}\,V_{a_2}\rangle=\lim_{a\rightarrow 0}\,\langle V_{a_1}\,V_{a_2}\,V_a\rangle,
\ee
or in other words, from defining the identity as the limit $\lim_{a\rightarrow 0} V_a$, with unit coefficient. This gives a 2-point function of the form \eqref{2pntfn-gen} with the spacelike analogous functions.

It is worth noticing that the timelike 2-point function \eqref{2pnt-fn-timelike} cannot be defined by an analogous limit from the timelike 3-point function \eqref{TL-3pnt-fn}. Indeed, the limit of vanishing charge, $\lim_{\alpha\rightarrow 0} V_\alpha$ does not yield the identity operator, but a non-degenerate primary of vanishing conformal dimension $V_0\neq V_{\langle 0\rangle}$ \cite{Harlow:2011ny,Ribault:2015sxa,Ikhlef:2015eua}.\footnote{This is a significant difference between the timelike and the spacelike theories, as in the latter such an operator does not exist because unitarity implies the identity is the only operator with vanishing dimension.} As a consequence, the limit $\lim_{\alpha\rightarrow 0}    C(\alpha_1,\alpha_2,\alpha)$ does not yield a diagonal expression-- i.e., a factor $\delta(\alpha_1-\alpha_2)$, and simply corresponds to the 3-point function with a primary of vanishing dimension. Nevertheless, the timelike 2- and the 3-point functions do happen to be related as
\be\label{reflection-coef-3pntfn}
\R(\alpha)=-2\beta\,   C(\alpha,\alpha,0).
\ee
The relative factor raises no issue since, as just explained, these two quantities need not be related, and eventually comes from the normalization of the 3-point structure constant. 

Besides following from the analytic continuation of the spacelike 2-point function, the timelike 2-point function can also be obtained from shift equations analogous to those for the 3-point structure constant. These follow from imposing crossing symmetry of a 4-point function where now two (instead of one) of the four insertions is either one of the two level-2 degenerate fields,  $V_{\langle1,2\rangle}$ or  $V_{\langle2,1\rangle}$; we review the derivation in appendix \ref{app-bulk-correls}. The resulting shift equations fix the $\Gamma$-function factors in \eqref{reflection-coef-timelike}, but again not its normalization, this eventually depends on the choice of operator normalization that also determines the 3-point structure constant.

\section{Boundary timelike Liouville}\label{sec:timelike boundary Liouville}

The study of boundary conformal field theory (BCFT) was pioneered by Cardy \cite{Cardy:1984bb,Cardy:1986gw,Cardy:1989ir} in the mid to late 80's.  Solving a BCFT amounts to determine the bulk and boundary primaries, and its lowest-point correlators, from which higher-point ones follow by factorisation.
Concretely, the essential data are the bulk 1-point function, the boundary and the bulk-boundary  2-point functions, and the boundary 3-point function.

The study of boundary states in spacelike Liouville theory was initiated by Fateev, Zamolodchikov and Zamolodchikov, and simultaneously by Teschner, in the early 2000s. 
Two boundary conditions emerged from these works: the FZZT \cite{Fateev:2000ik,Teschner:2000md} and the ZZ solutions \cite{Zamolodchikov:2001ah}. In this section, we derive the bulk 1-point and the boundary 2-point structure constants of timelike Liouville theory on the upper half plane, with boundary conditions analogous to those of the spacelike FZZT. 

In spacelike Liouville, the bulk-boundary 2-point function reduces to the bulk 1-point function  when the boundary operator's charge is taken to zero \cite{Hosomichi:2001xc}, since in that case this operator becomes the identity. However, in timelike Liouville this need not be the case, since at least in the bulk spectrum, the vanishing charge limit does not imply the operator becomes identity, as explained in section \ref{section:TL}. 

\vspace{5mm}

The action for timelike Liouville theory on a space with a boundary is
\be
S_{BtL}[\chi]=\frac{1}{4\pi}\int\limits_{\mathcal M} d^2x \sqrt{h}\left(-(\nabla\chi)^2-q R_h\chi+4\pi\mu\,e^{2\b\chi}\right)+\frac{1}{2\pi}\int\limits_{\partial\mathcal M}
d\xi \,h^{1/4}\,\left(-q\,K_{h} \chi +2\pi\mu_B\,e^{\b\chi}\right),
\ee
where $\xi$ is the coordinate of the boundary, $K_h$ is the boundary curvature, and $\mu_B$ is the boundary cosmological constant, which parametrises the boundary condition.
The boundary terms are fixed by demanding conformal invariance of the boundary condition.

We consider the fiducial metric $h$  to be the flat disk. This is equivalent to the upper half-plane, $ds^2=dz\,d\bar z$ with $\text{Im}\,z>0$, as long as the boundary condition on the field at infinity is taken to be
\be\label{asymptotic-field-timelike-halfplane}
\chi(z)\xlongrightarrow{|z|\rightarrow\infty} -2\,q\log |z| +\mathcal O(1).
\ee
The action on the upper half-plane becomes
\be
S_{BtL}[\chi]=\frac{1}{2\pi}\int\limits_{\text{Im}\,z>0} dz\,d\bar z \,\left(- \partial\chi\bar\partial\chi+\pi\mu\,e^{2\b\chi}\right)+\int\limits_{-\infty}^\infty dx\,\mu_B\,e^{\b\chi},
\ee
where $x$ runs now over the real axis. The background charge no longer appears in the action, and is instead introduced through the asymptotics of the field \eqref{asymptotic-field-timelike-halfplane}.

Similar to the bulk primaries $V_\a=e^{-2\a\chi}$, the boundary primary operators are given by
\be
B_\d=e^{-\d\chi},
\ee
with conformal dimension $\Delta_\d=\d(\d-q)$.

\subsection{Bulk 1-point function}\label{sec:1-pnt-fn}

The bulk one-point function for any CFT on the upper half plane  is
\begin{equation}\label{disk_1pt-timelike}
\langle V_\a(z)\rangle = \frac{  U(\a)}{|z-\bar{z}|^{2\Delta_\a}}.
\end{equation}
Besides depending on the charge $\a$ of the operator, the 1-point structure constant $U(\a)$ must further depend on the cosmological constant $\mu_B$ which parametrises the boundary condition; here we leave this dependence implicit to not clutter the notation. 

To determine the structure constant $  U(\a)$, it is convenient to follow the bootstrap approach and derive shift relations \cite{Teschner:1995yf} similar to those reviewed in section \ref{correlators} and appendix \ref{app-bulk-correls} for  2- and  3-point structure constants.
This was done originally in \cite{Fateev:2000ik} in the spacelike Liouville regime, but we will closely follow the more normalization-explicit approach presented in \cite{Ribault:pohang_notes}.

\begin{figure}[h]
	\centering
	\includegraphics[scale=0.5]{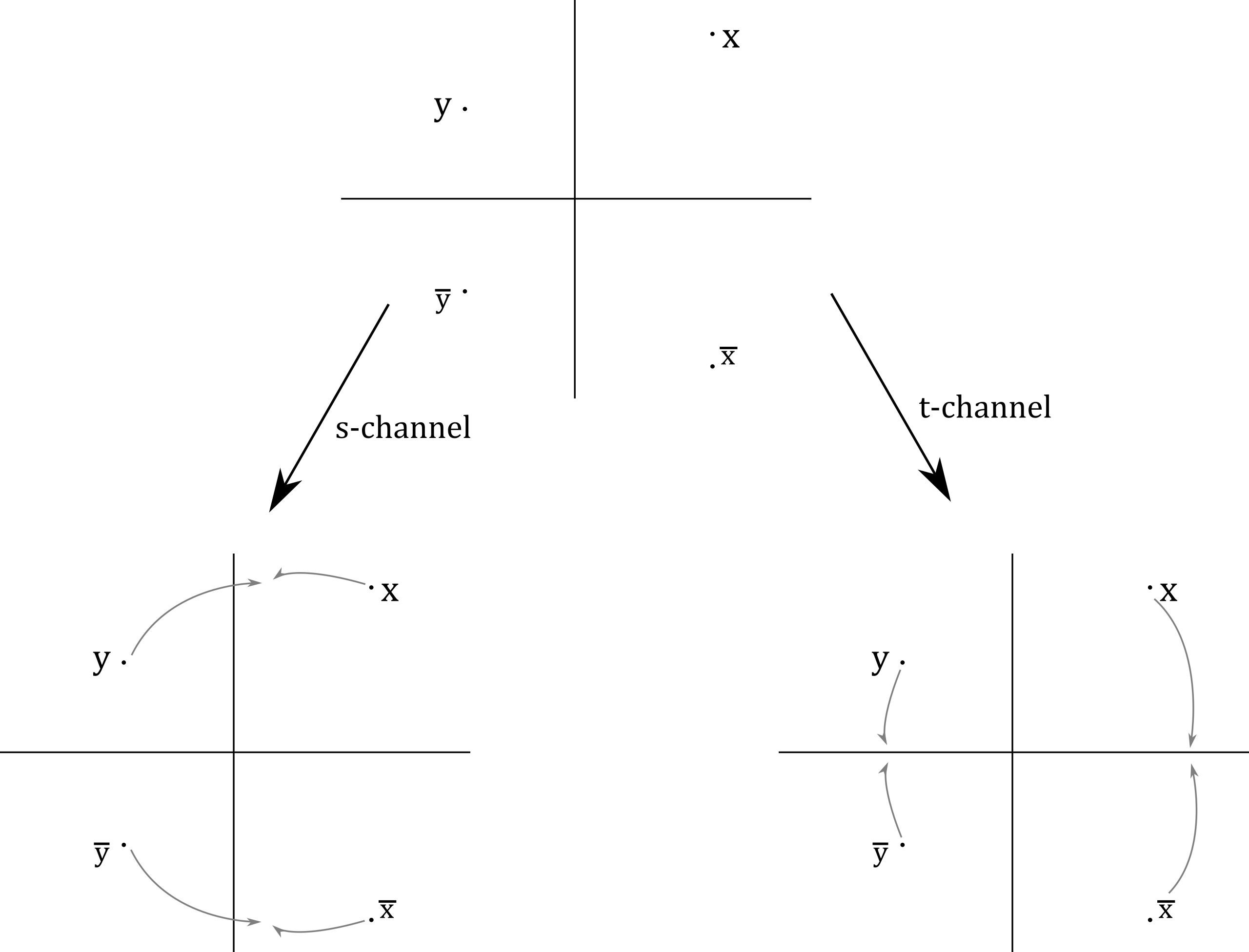}
	\caption{s- and t- channel decomposition for two points on the upper half plane.}
	\label{fig:st_chan}
\end{figure}

As explained in \ref{correlators}, to derive shift equations for the 3-point structure constant, we consider 4-point functions with one insertion of either one of the two level-2 degenerate operators.
Similarly, to derive shift equations for the bulk 1-point structure constant, we  consider 2-point functions with an insertion of either one of the two level-2 degenerate operators. These two operators are
\be\label{degenerate-ops-timelike}
V_{\langle1,2\rangle}\,:\,\, \a_{1,2}=\frac{\beta}{2},\quad\Delta_{1,2}=-\frac{1}{2}+\frac{3}{4}\beta^2,\qquad\quad
V_{\langle2,1\rangle}\,:\,\, \a_{2,1}=-\frac{1}{2\beta},\quad\Delta_{2,1}=-\frac{1}{2}+\frac{3}{4\beta^2}.
\ee
We start by looking at $\langle V_\a(x)V_{\langle1,2\rangle}(y)\rangle$, where $x$ and $y$ belong to the upper half plane. 
The corresponding cross-ratio is 
\be\label{cross-ratio}
z = \frac{(y-x)(\bar{y}-\bar{x})}{(y-\bar{x})(\bar{y}-x)}.
\ee
Two OPEs are now possible, see figure \ref{fig:st_chan}: one where the two operators are brought close together, $x\rightarrow y,\bar{x}\rightarrow \bar{y}$, hence $ z\rightarrow 0$, and one where the operators are brought close to the boundary $x\rightarrow \bar{x}, y\rightarrow \bar{y}$, hence $z\rightarrow 1$. The kinematical part of each of these OPEs can be seen to correspond respectively to the s- and t-channel degenerate 4-point conformal blocks $ \F^s_\pm(z)$ and $ \F^t_\pm(z)$, given in  \eqref{conf-bloc-sch-timelike} and \eqref{conf-bloc-tch-timelike}, where one of the three generic insertions is in this case taken to be the same degenerate $V_{\langle1,2\rangle}$.

Concretely, the s-channel decomposition reads
\be\label{s-channel decomp-1pntfn-TL}
\langle V_\a(x)\,V_{\langle1,2\rangle}(y)\rangle=\sum\limits_\pm  C_\pm(\a)\,  U\left(\a\pm\tfrac{\beta}{2}\right)\, {\mathcal F}
^s_\pm(z),
\ee
where we have omitted length pre-factors  $\sim|x-y|$ for simplicity.  $  C_{\pm}(\a)$ are the bulk OPE coefficients
\be\label{bulk-OPE-timelike}
V_\a\,V_{\langle 1,2\rangle}\sim    C_+(\a)\,V_{\a+\frac{\b}{2}}+    C_-(\a)\,V_{\a-\frac{\b}{2}},
\ee
and the conformal blocks ${\mathcal F}
^s_\pm(z)$ are evaluated on $\a_1=\a_3=\a$ and $\a_2=\a_{1,2}$.

The t-channel decomposition requires instead the bulk-boundary OPEs, which for the case of the degenerate operator is
\be\label{bB-OPE-timelike}
V_{\langle 1,2\rangle}\sim   c_-\,B_{\langle 1,1\rangle}+  c_+\,B_{\langle 1,3\rangle}.
\ee
The operators on the right-hand side are boundary degenerate operators $B_{\langle m,n\rangle}$, with $\d_{m,n}=\frac{1-m}{2\beta}-\frac{(1-n)\b}{2}$. $B_{\langle 1,1\rangle}$ corresponds to the boundary identity, with $\d_{1,1}=0$. The bulk-boundary OPE coefficients depend on the boundary cosmological constant $\mu_B$.

\begin{figure}[h]
	\centering
	\includegraphics[scale=0.7]{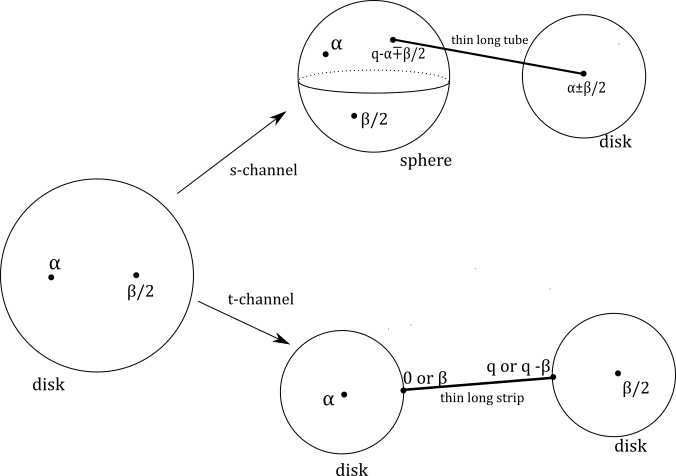}
	\caption{Bunching the two operators on the disk together is like pinching off a sphere with the two operators, while taking them far apart is like decomposing the disk into two disks.}
	\label{fig:disk_dec}
\end{figure}

 The t-channel decomposition then becomes
\be\label{t-channel decomp-1pntfn-TL}
\langle V_\a(x)\,V_{\langle1,2\rangle}(y)\rangle=   c_-\,  U(\a)\, {\mathcal F}^t_-(z) +   c_+\,  R(\a,\d_{1,3})\, {\mathcal F}^t_+(z),
\ee
where again the conformal blocks are evaluated on $\a_1=\a_3=\a$ and $\a_2=\a_{1,2}$. $  R(\a,\d)$ is the structure constant corresponding to the bulk-boundary 2-point function $\langle V_\a\, B_\d\rangle$ on the disk, and  the 1-point structure constant comes from the bulk-boundary 2-point function $R(\a,0)=  U(\a)$, see figure \ref{fig:disk_dec}. 
We will not need the expression for the structure constant $R(\a,\d)$ to derive the shift equations.

Crossing symmetry, or in other words associativity of the OPE, implies the s- and t-channel decompositions have to equal each other, hence
\be
  C_+(\a)\,  U\left(\a+\tfrac{\beta}{2}\right)\, {\mathcal F}
^s_+(z)+  C_-(\a)\,  U\left(\a-\tfrac{\beta}{2}\right)\, {\mathcal F}
^s_-(z)=  c_-\,  U(\a)\, {\mathcal F}^t_-(z) +  c_+\,  R(\a,\a_{1,3})\, {\mathcal F}^t_+(z).
\ee
 Using the relation between the s- and t-channel conformal blocks $ \F^s_i(z) = \sum_{j=\pm}\mathcal{B}_{ij}\, \F^t_j(z)$, with the elements of the degenerate fusing matrix $\mathcal B_{ij}$ given by \eqref{fusing-matrix-elements} and \eqref{ABC} and evaluated at $\a_1=\a_3=\a, \,\a_2=\a_{1,2}$, we obtain the shift equation
\be\label{shift-eqn-1pntfn-TL-OPEs}
c_-\,  U(\a)=  C_+(\a)\,  U\left(\a+\tfrac{\beta}{2}\right)\,\mathcal B_{+-}+  C_-(\a)\,  U\left(\a-\tfrac{\beta}{2}\right)\,\mathcal B_{--}.
\ee
At this point, to make any progress with this equation we need to input the expressions for the (bulk) OPE coefficients, and hence we need to partially fix our normalization, namely the $\alpha$-dependent part. 
	Our choice for such normalization has been explained in detail in appendix \ref{app:DF-integrals}, and consists of identifying Liouville correlators  on the complex plane whose charges satisfy $\a_t=q+n\b$ with $n$ a non-negative integer, to their expression obtained from a perturbative computation. They then become identified with integrated correlators of the free theory with appropriate insertions of screening operators.

With such a normalization, one realises that  the first OPE coefficient $  C_+(\a)$ is in fact an $\a$-independent quantity, and fixing it amounts to fixing the $\alpha$-independent normalization of the 3-point structure constant. For now, we keep denoting it as $  C_+$. The second OPE coefficient follows from the integral of a free correlator with insertion of a screening operator (see \eqref{C-OPE-coef-screening}). The ratio of the two coefficients is
\be\label{ratio-bulkOPEs-spacelike}
\frac{C_-(\a)}{C_+}=-\frac{\pi\mu}{\gamma(\b^2)}\frac{\gamma(2\a\b+\b^2-1)}{\gamma(2\a\b)}.
\ee

We require the structure constant $U(\a)$ to satisfy the reflection property $U(\a) = \mathcal{R}(\a)U(q-\a)$ where $\mathcal{R}(\a)$ is given in \eqref{reflection-coef-timelike}. Following \cite{Ribault:pohang_notes} we define
\be\label{definition_A}
\mathcal A(\a) :=\left[\pi\,\mu\,\gamma(-\beta^2)\right]^{\frac{\alpha}{\beta}}\,\,(2\a-q)\,\,\Gamma\left(\b^{-1}(q-2\a)\right)\,\Gamma\left(\beta(2\a-q)\right),
\ee
so that
$
\R(\a)=-\frac{  \mathcal A(\a)}{  \mathcal A(q-\a)}.
$ Now, in terms of

\be
U_{\sub{R}}(\a):=\frac{  U(\a)}{  \mathcal A(\a)},
\ee
the reflection property takes a particularly simple form: 
\be\label{reflection-pseudoinv}
U_{\sub{R}}(\a)=-U_{\sub{R}}(q-\a).
\ee
The shift equations too take a very simple form in terms of $U_{\sub{R}}(\a)$.
Introducing the expressions for the OPE coefficients, the degenerate fusing matrix elements, and the reflection-invariant 1-point function into \eqref{shift-eqn-1pntfn-TL-OPEs}, the shift equation becomes
\be\label{shift-eqn-1pntfn-timelike}
2\,\cosh(\pi\beta s) \,  U_{\sub{R}}(\a)=  U_{\sub{R}}\left(\a+\tfrac{\beta}{2}\right)+  U_{\sub{R}}\left(\a-\tfrac{\b}{2}\right),
\ee
where a new parameter $s$, satisfying
\be\label{s-param-timelike-OPE}
2\,\cosh(\pi\beta s) =-\frac{  c_-}{  C_+}\frac{1}{\b^2\,\sqrt{\pi\mu\gamma(-\b^2)}}\frac{\Gamma(\b^2)}{\Gamma(-1+2\b^2)},
\ee
has been introduced for reasons to become clear just below. This parameter depends on $\mu_B$ through the bulk-boundary OPE coefficient $c_-$.

An analogous equation can be derived from crossing symmetry of the 2-point function with now the other level-2 degenerate operator, $\langle V_\a(x)\,V_{\langle2,1\rangle}(y)\rangle$. Keeping in mind the Liouville duality
\begin{align}\label{duality-timelike}
&\b\rightarrow-\frac{1}{\b},\\
&\mu\rightarrow\tilde \mu, \quad\text{with}\quad \pi\mu\gamma(-\b^2)=\left[\pi\,\tilde\mu\,\gamma(-1/\b^2)\right]^{-\b^2},
\end{align}
this second equation reads
\be\label{shift-eqn-1pntfn-timelike-dual}
2\,\cosh\left(\frac{\pi s}{\b}\right)\,  U_{\sub R}(\a)=  U_{\sub R}\left(\a-\tfrac{1}{2\b}\right)+  U_{\sub R}\left(\a+\tfrac{1}{2\b}\right),
\ee
where the parameter $s$ further satisfies
\be\label{s-param-timelike-OPE-dual}
2\,\cosh\left(\frac{\pi s}{\b}\right)=-\frac{\tilde{  c}_-}{\tilde{  C}_+}\frac{\b^2}{\sqrt{\pi\tilde\mu\gamma(-1/\b^2)}}\frac{\Gamma(1/\b^2)}{\Gamma(-1+2/\b^2)},
\ee
which includes in this case the bulk and bulk-boundary
 OPE coefficients involving $V_{\langle2,1\rangle}$, $\tilde{ C}_+$ and  $\tilde{c}_-$  respectively.

The two shift equations obtained for $U_{\sub R}(\a)$ \eqref{shift-eqn-1pntfn-timelike} and \eqref{shift-eqn-1pntfn-timelike-dual}, admit as solutions any linear combination of $U_{\sub R}(\a)=e^{\pm 2\pi s \a}$.
However, we must further impose 
$U_{\sub R}(\a)=-U_{\sub R}(q-\a).
$
This restricts the relative coefficient between the two solutions to be $-e^{2\pi s q}$, or in other words, restricts the general solution to be proportional to the combination 
\be
e^{\pi s(2\a-q)}-e^{-\pi s (2\a-q)}\sim \sinh(\pi s(2\a-q)).
\ee
The 1-point structure constant then becomes
\be\label{unnormalised-1-pnt-fn-timelike}
  U(\a)= C\,\left[\pi \mu \gamma(-\b^2)\right]^{\frac{\a}{\b}}\, (2\a-q)\,\Gamma\left(\b^{-1}(q-2\a)\right)\Gamma\left(\b(2\a-q)\right)\sinh\left(\pi s (2\a - q)\right),
\ee
where the $\a$-independent normalization $C$ remains unfixed by the shift equations. 

\subsection*{Normalization}

	We now proceed to fix the $\a$-independent normalization of the 1-point structure constant. We again resort to the perturbative method explained in appendix  
\ref{app:DF-integrals}. Concretely, we identify Liouville correlators \textit{on the upper half-plane} whose charges satisfy $2\a_t+\d_t=q+n\b$ with $n$ a non-negative integer, to integrated correlators of the free theory with additional screening operators, as given by \eqref{Liouv-corr-Cgas-corr-boundary}. 
Using such method, a 1-point function with charge $2\a=q+n \b$ is such that
\be
\langle V_\a(z)\rangle= \frac{1}{2\beta}\sum_{k=0}^{\lfloor n/2 \rfloor}\frac{(-\mu)^k(-\mu_B)^{n-2k}}{k!(n-2k)!} \Big\langle V_{\a}(z)\left(\int_{\sub\text{Im}\,z>0}d^2z_i\,V_{-\b}(z_i)\right)^k\left( \int_{-\infty}^{\infty} dx_j\,B_{-\b}(x_j)\right)^{n-2k} \Big\rangle_{\sub 0},
\ee
where the correlator on the right-hand side is evaluated on the timelike free theory on the upper half plane, and is hence given by \eqref{free-correl-bdry}.

In particular, in the case $\a=q/2$, hence $n=0$, 
\be\label{normalization-1pnt-fn}
\langle V_{q/2}(z)\rangle= \frac{1}{2\beta}\langle V_{q/2}(z)\rangle_{\sub 0}= \frac{1}{2\beta}|z-\bar z|^{q^2/2}.
\ee
Back to our expression for the 1-point function \eqref{unnormalised-1-pnt-fn-timelike}, we hence demand
\be\label{normalization-1pnt-fn-T}
  \lim\limits_{\a\rightarrow q/2}U(\a)= \frac{1}{2\beta},
\ee
which fixes the structure constant to be
\small\begin{equation}\label{1-pnt-fn-normalised}
	 {U}(\a) = \frac{1}{2\beta}\left[\pi \mu \gamma(-\b^2)\right]^{\frac{2\a-q}{2\b}} \Gamma\left(\b^{-1}(q-2\a)+1\right)\Gamma\left(\b(2\a-q)+1\right)\frac{\sinh\left(\pi s (2\a - q)\right)}{\pi s (2\a - q)}.
\end{equation}\normalsize

To determine the $\mu_B$-dependence of $s$, notice that the bulk-boundary OPE coefficient $c_-$ is related to the bulk-boundary 2-point structure constant as
\be
c_-=\frac{R(\a_{1,2},\d_{1,1})}{D(\d_{1,1})}=R(\a_{1,2},q-\d_{1,1})=R(\b/2,q),
\ee
where $D(\d)$ is the boundary 2-point structure constant, to be defined and computed in the next section \ref{sec:bdry-2pnt-fn}. The only property of this 2-point function we have used here is that it acts as the boundary reflection coefficient, just as for the bulk 2-point, such that $B_\d=D(\d)\,B_{q-\d}$.  

The charges in the corresponding 2-point function $\langle V_{\langle 1,2\rangle} \,B_{q}\rangle$  are such that $2\a_t+\d_t=\b+q$, i.e. $n=1$, so this 2-point function admits the perturbative expression \eqref{Liouv-corr-Cgas-corr-boundary}:
\be
\langle V_{\langle1,2\rangle}(z)\,B_{q}(x)\rangle=-\frac{\mu_B}{2\beta}\,\Big\langle V_{\langle1,2\rangle}(z)\,B_q(x)\int_{-\infty}^\infty dy\,B_{-\b}(y)\Big\rangle_{\sub 0}.
\ee
Its structure constant can then be obtained by fixing the two unintegrated insertions,
	\begin{equation}
\frac{R(\b/2,q)}{2^{2\Delta_{1,2}}}=-\frac{\mu_B}{2\beta}\,\,\int_{-\infty}^\infty\left\langle V_{\langle 1,2\rangle}(i)\,B_q(\infty)\,B_{-\b}(y) \right\rangle_{\sub 0} dy =- \frac{2^{\b^2/2}\,\mu_B}{2\b}\,\,\int_{-\infty}^\infty \frac{dy}{|i-y|^{2\b^2}},
\end{equation}
where in the second step we have used the expression for the free correlators on the upper half plane \eqref{free-correl-bdry}. The factor of $2^{2\Delta_{1,2}}$ accounts for the fact that the correlator is evaluated at $z=i$. 
Finally, the bulk-boundary OPE coefficient results in
\begin{equation}\label{eq:special_bulkbdry_SL}
c_-=R(\b/2,q) = -\frac{\pi\mu_B}{\b}\,\frac{\Gamma(-1+2\b^2)}{\Gamma^2(\b^2)}.
\end{equation}
Substituting this expression into the definition of the $s$ parameter \eqref{s-param-timelike-OPE}, and further using that with our normalization for the bulk correlators $C_+=-1/2\beta$ (see appendix \ref{app:DF-integrals}), we obtain
%
	\begin{equation}\label{eq:s_muB_TL}
\cosh (\pi \b s) = -\mu_B\,\sqrt{\frac{\sin(\pi \b^2)}{-\mu}},
\end{equation}
which gives the explicit dependence of $s$ on the boundary cosmological constant. 

The Liouville duality \eqref{duality-timelike} needs then to be complemented by $\mu_B\rightarrow\tilde\mu_B$ such that $s$ defined as above further satisfies
\begin{equation}\label{eq:s_muB_TL_dual}
\cosh\left(\frac{\pi s}{\b}\right) = -\tilde{\mu}_B\sqrt{\frac{\sin\left(\pi/ \b^{2}\right)}{-\tilde{\mu}}}.
\end{equation}
Assuming $\b,\mu,\mu_B$ and their duals $\in\mathbb R$, and given that we want the semiclassical limit at $\b\rightarrow0$, equation \eqref{eq:s_muB_TL} determines different reality conditions for $s$ depending on the signs of the cosmological constants. Concretely, $\mu>0$ requires $s\in \mathbb C$ with an imaginary part proportional to $1/\b$. On the other hand, $\mu<0$ requires $s$ purely imaginary if the r.h.s of \eqref{eq:s_muB_TL} is $<1$. If instead it is $>1$, then $s$ can be real for $\mu_B<0$, or else must be complex again with imaginary part proportional to $1/\b$. 

The analogous conclusions can be drawn from the dual equation \eqref{eq:s_muB_TL_dual}, depending on the signs of the dual cosmological constants. In the cases when $s$ is required to be complex, the imaginary part is now fixed to be proportional to $\b$ instead. Since $s$ needs to satisfy both equations, these two conditions restrict the ranges of the constants for which the 1-point function \eqref{1-pnt-fn-normalised} is a valid solution. In particular, when $\mu,\tilde\mu>0$, the 1-point function is only a solution for certain rational values of $\b^2$. Instead, having all the cosmological constants negative seems to be allowed for generic value of $\b$ or the central charge. A more thorough analysis of all the cases would require checking the compatibility of the signs of the cosmological constants as established by their duality relations, and depending on the values of $\beta$, which we leave for future work.

%
%
%

Finally, notice that the 1-point structure constant \eqref{1-pnt-fn-normalised} is invariant under $s\rightarrow-s$, consistent with the above relation \eqref{eq:s_muB_TL} and \eqref{eq:s_muB_TL_dual} being insensitive to the sign of $s$. 

\subsection{Boundary 2-point function}\label{sec:bdry-2pnt-fn}

\begin{figure}[h]
	\centering \includegraphics[scale=0.6]{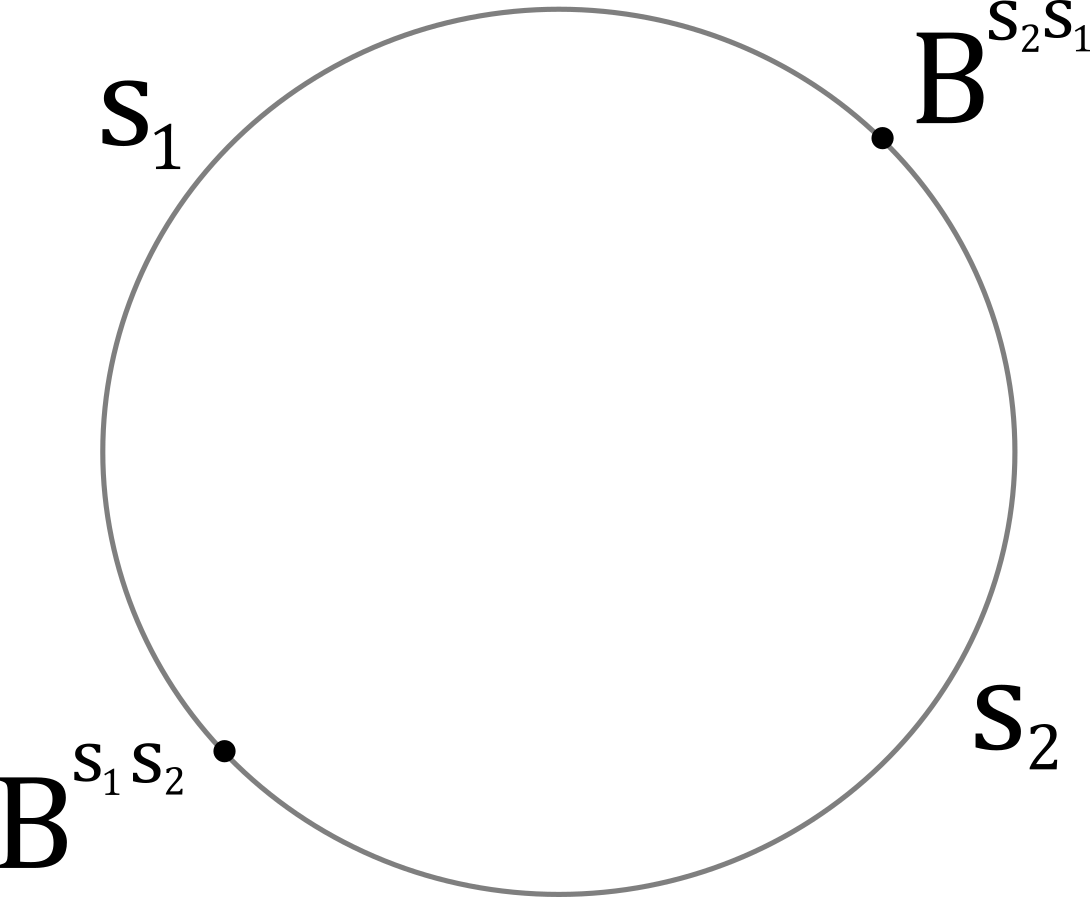}
	\caption{A disk model of the boundary 2-point function}
	\label{fig:disk_2pt}
\end{figure}

	On the upper half plane, the boundary 2-point function is given by
\begin{equation}
\left\langle B_{\d_1}^{\mu_1\mu_2}(x)B_{\d_2}^{\mu_2\mu_1}(0)\right\rangle =\frac{D(\d_1|\mu_1,\mu_2)\,\delta(\d_1-\d_2)+\delta(q-\d_1-\d_1)}{|x|^{\Delta_{1}+\Delta_{2}}}
\end{equation}
where $\mu_1$,$\mu_2$ are the boundary cosmological constants on either side of the operator $B_\d^{\mu_1\mu_2}$, and $x$ takes values on the real line. The factor of unit in front of the second Dirac delta in the numerator means that the boundary 2-point structure constant acts as a boundary reflection coefficient such that
\be\label{reflection-rel-bdry}
B_\d^{\mu_1 \mu_2}=D(\d|\mu_1,\mu_2)\,B_{q-\d}^{\mu_1\mu_2}.
\ee
 Using the parametrization \eqref{eq:s_muB_TL}, we may also denote this operator as $B_\d^{s_1s_2}$, and the structure constant as $D(\d|s_1,s_2)$. See figure \ref{fig:disk_2pt} for a depiction of this correlator. The goal of this subsection is to  compute $D(\d|s_1,s_2)$.

 Consider the boundary 3-point function $\langle B_\d^{s_1s_2}B_{\langle1,3\rangle}^{s_2s_2}B_{\d-\b}^{s_2s_1}\rangle$ with the level-3 degenerate operator\footnote{We refer the reader to \cite{Fateev:2000ik} for an explanation as to why using a level-2 degenerate field as we have been doing for bulk fields does not work in this case.} 
 \be
B_{\langle1,3\rangle}^{ss} \,:\,\, \d_{1,3}=\b.
 \ee
 
\begin{figure}
	\centering \includegraphics[scale=0.7]{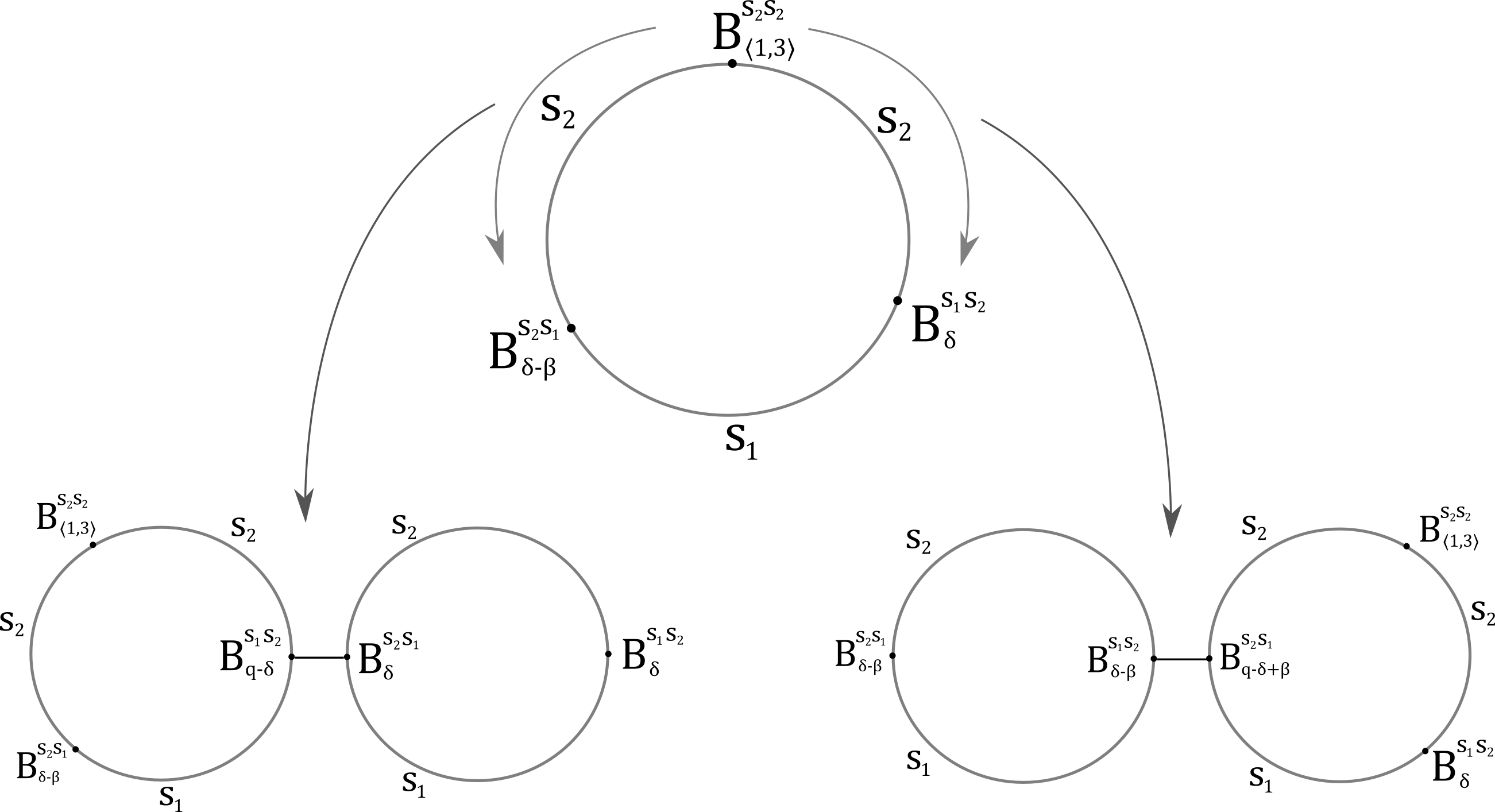}
	\caption{Two OPEs of the 3-point function with the degenerate operator $B_{\langle1,3\rangle}$.}
	\label{fig:disk_2pt_2}
\end{figure}

  Two OPEs arise corresponding to the degenerate operator approaching either one of the other two operators in the correlator. Taking $B_{\langle1,3\rangle}^{s_2s_2}$ close to $B_\d^{s_1s_2}$ gives the OPE
  \begin{equation}
  \begin{aligned}
B_\d^{s_1s_2}B_{\langle1,3\rangle}^{s_2s_2}\sim  c_+(\d)\,B_{\d+\b}^{s_1s_2}+c_0(\d)\,B_{\d}^{s_1s_2}+c_-(\d)\,B_{\d-\b}^{s_1s_2},
  \end{aligned}\end{equation}
  where the boundary OPE coefficients\footnote{These boundary OPE coefficients $c_\sigma$ are not to be confused with the bulk-boundary ones in \eqref{bB-OPE-timelike}.} satisfy
  \be
  c_\sigma(\d) =\frac{\langle B_\d^{s_1s_2}B_{\langle1,3\rangle}^{s_2s_2}B_{\d+\sigma \b}^{s_2s_1}\rangle}{D(\d+\sigma \b|s_1,s_2)}
  = \langle B_\d^{s_1s_2}B_{\langle1,3\rangle}^{s_2s_2}B_{q - \d-\sigma \b}^{s_2s_1}\rangle
  \ee
   with $\sigma\in\{+,0,-\}$. In the second step we have used the fact that the boundary 2-point structure constant acts as a reflection coefficient as in \eqref{reflection-rel-bdry}.
  With this OPE, the 3-point function becomes
\begin{equation}
\begin{aligned}
\left\langle B_\d^{s_1s_2}B_{\langle1,3\rangle}^{s_2s_2}B_{\d-\b}^{s_2s_1}\right\rangle 
= c_-(\d)\,D(\d-\b|s_1,s_2).
\end{aligned}\end{equation}
Instead, taking $B_{\langle1,3\rangle}^{s_2s_2}$ to $B_{\d-\b}^{s_2s_1}$ gives
\begin{equation}
\begin{aligned}
\left\langle B_\d^{s_1s_2}B_{\langle1,3\rangle}^{s_2s_2}B_{\d-\b}^{s_2s_1}\right\rangle 
= c_+(\d-\b)\,D(\d|s_1,s_2)
\end{aligned}\end{equation}
where 
$c_+(\d-\b) = \langle B_{q - \d}^{s_1s_2} B_{\langle1,3\rangle}^{s_2s_2}B_{\d-\b}^{s_2s_1}\rangle$.
See figure \ref{fig:disk_2pt_2} for a representation of the two OPEs.

Equating both expansions leads to the shift relation
\begin{equation}\label{eq:bdry_2pt_shift_TL}
\frac{D(\d|s_1,s_2)}{D(\d-\b|s_1,s_2)} = \frac{c_-(\d)}{c_+(\d-\b)}.
\end{equation}

As was required for the shift equation of the  bulk 1-point structure constant, we now need to fix the ratio of OPE coefficients. 
	As before, we do this via perturbative screening integrals. Namely, we compute each of the two coefficients separately using the expression for Liouville correlators in terms of integrated free correlators \eqref{Liouv-corr-Cgas-corr-boundary} and \eqref{free-correl-bdry}. Their $\delta$-independent normalization is not important since we require only their ratio.

The coefficient $c_+(\d-\b)$ needs no screening operators, since the momenta of the operators add to $q$. On the other hand, the momenta of $c_-(\d)$ add to $q+2\b$, which can be screened by either a single bulk field $V_{-\b}$ or two boundary fields $B_{-\b}$, leading to two contributions. The required integrals are evaluated in \cite{Fateev:2000ik}, and give
	\small\begin{equation}
\begin{aligned}
\frac{D(\d|s_1,s_2)}{D(\d-\b|s_1,s_2)}=& \frac{4\mu \,\b^4\,\gamma(-\b^2)}{\pi}\,
\Gamma(2\d\b-1)\,\Gamma(1-2\d\b)\,\Gamma(2\d\b-1+\b^2)\,\Gamma(1-2\d\b+\b^2)\\
&\times\, \sin\left[\pi \b\left(\d+ i\tfrac{s_1+s_2}{2}\right)\right]
\sin\left[\pi \b\left(\d- i\tfrac{s_1+s_2}{2}\right)\right]
\sin\left[\pi \b\left(\d+ i\tfrac{s_1-s_2}{2}\right)\right]
\sin\left[\pi \b\left(\d- i\tfrac{s_1-s_2}{2}\right)\right].
\end{aligned}
\end{equation}\normalsize
Using the shift relations \eqref{eq:Gamma_b_shift} and \eqref{eq:double_sin_shifts}, and additionally requiring that $D(\d|s_1,s_2)D(q-\d|s_1,s_2)=1$, it follows that
	\begin{equation}\label{bdry2pt_TL}
	\begin{aligned}
		 {D}(\d|s_1,s_2) = \left[\pi\mu\gamma(-\b^2)\b^{2+2\b^2}\right]^{(2\d-q)/2\b}\frac{\Gamma_\b(q-2\d+\b)}{\Gamma_\b(2\d-q+\b)}\prod_{\pm\pm'}S_\b(\d+\b\pm i\tfrac{s_1\pm' s_2}{2})
	\end{aligned}
\end{equation}
where the $\pm\pm' $ indicates the two sets of signs have to be taken to be independent of each other, so that the product consists of four factors. $\Gamma_\b$ is the double Gamma function \eqref{eq:Gamma_b_def}, and $S_\b$ is the double-sine function \eqref{double-sine-fn}, defined by a ratio of double Gamma functions (see appendix \ref{app:special functions} for more properties of these functions).
The overall sign in this expression is such that $ {D}(q/2|s_1,s_2)=1$, just as the bulk reflection coefficient satisfies  $\R(q/2)=1$.\footnote{This is analogous to the spacelike case, where the boundary 2-point structure constant of the operator with charge $Q/2$ is $-1$, just as the spacelike reflection coefficient is $-1$.}

\section{Discussion}\label{sec:discussion}

In this work, we have studied boundary timelike Liouville theory on the Euclidean disk or upper half plane, by computing two of the basic boundary CFT data: the bulk 1-point  and the boundary 2-point structure constants. Similar to the FZZT solutions for boundary Liouville theory in the spacelike regime \cite{Fateev:2000ik,Teschner:2000md}, we find a family of conformal boundary conditions parametrised by the boundary cosmological constant $\mu_B$.
The bulk 1-point structure constant, given in \eqref{1-pnt-fn-normalised}, exhibits a $\sinh$ behaviour of the operator charge, while the boundary 2-point function \eqref{bdry2pt_TL} exhibits a product of double-sine  functions  of the charge.

It is instructive to compare our results to the  spacelike Liouville expressions  for the same objects.
In spacelike Liouville theory, the bulk 1-point and the boundary 2-point structure constants were determined \cite{Fateev:2000ik,Teschner:2000md} with the same bootstrap approach as we have employed in this work. Namely, crossing symmetry of correlators with spacelike degenerate operator insertions was used to derive shift equations for the structure constants. The shift equations we have derived in the timelike regime coincide with the  analytic continuation  $q=-iQ$, $\b=ib$, $\a=-ia$ of the spacelike ones. 

 The solution for the 1-point structure constant  in the spacelike regime is the famous FZZT  1-point function \cite{Fateev:2000ik,Teschner:2000md}
\begin{equation}\label{FZZT}
	U_{\small\sub{\text{FZZT}}}(a)=2\,\left[\pi \mu \gamma(b^2)\right]^{\frac{Q-2a}{2b}}\Gamma\left(b^{-1}(2a-Q)+1\right)\Gamma\left(b(2a-Q)+1\right)\frac{\cosh\left[\pi s (2a - Q)\right]}{(2a-Q)},
\end{equation}
where the parameter $s$ is in this case defined by
\begin{equation}\label{eq:s_muB_SL}
	\cosh (\pi b s) = \mu_B \,\sqrt{\frac{\sin(\pi b^2)}{\mu}}.
\end{equation}
Comparing with our 1-point function solution for the timelike regime \eqref{1-pnt-fn-normalised}, the most important difference  is in the \textit{cosh} versus the \textit{sinh} dependence on the Liouville charge $\a$. Naively, it would seem that the timelike 1-point structure constant should be  the analytic continuation of the spacelike one, since the analytic continuation is well defined:\footnote{There might seem to be two different analytic continuations depending on whether $s$ is also analytically continued or not, leading to either $\cosh \left[\pi \sigma (2\a-q)\right]$ or  $\cos \left[\pi s(2\a-q)\right]$. However, notice that these two are the same solution, since this would also change the definition of $s $ in terms of $\mu_B$ accordingly.}
$
\cosh\left[\pi s (2a - Q)\right]\rightarrow\cosh \left[\pi \sigma (2\a-q)\right]
$, where $\sigma$ would be defined as $\sigma=i s$.
However, we are forced to discard this solution because it does not have the desired reflection property, as we explain next.

The best way to obtain the correct 1-point function is to look for the valid solutions to the shift equations  in each regime of the theory. Doing so, we realise that \textit{in both} the spacelike and the timelike regime, the shift equations admit \textit{both} the $\sinh$ and the $\cosh$ solution (and in fact any  linear combination of the two). It is then the reflection property of the 1-point function that selects either one or the other solution, and it happens to select a different answer in each regime. 

The reason why reflection determines a different solution in each regime is eventually encoded in the expression of the reflection coefficient. The spacelike reflection coefficient as determined from reflection of the DOZZ 3-point structure constant, reads
\begin{equation}\label{reflection-coef-spacelike}
	\mathcal{R}_{\sub{\text{DOZZ}}}	(a) = -\left[\pi \mu \gamma(b^2)\right]^{\frac{Q-2a}{b}} \frac{\Gamma\left(b^{-1}(2a-Q)\right)\Gamma\left(b(2a-Q)\right)}{\Gamma\left(b^{-1}(Q-2a)\right)\Gamma\left(b(Q-2a)\right)}.
\end{equation}
This is not exactly equal to the analytic continuation of the timelike reflection coefficient $\mathcal R(\a)$ given in \eqref{reflection-coef-timelike}: there is an additional overall minus sign. In other words, in each regime the corresponding 3-point structure constants determine  reflection coefficients which are related by analytic continuation up to a minus sign.
In the spacelike regime then, the 1-point function $U_{\sub R}$ satisfying the shift equations is exactly reflection invariant, while in the timelike case it is reflection invariant up to a minus sign, see equation \eqref{reflection-pseudoinv}. 

One may wonder whether the difference between the $\cosh$ and the $\sinh$ behaviours in each regime is a pure artifact of the normalization. On the one hand, we could define a \textit{normalization-invariant} 1-point structure constant $U_N(\alpha)\equiv U(\alpha)/N(\alpha)$, as done with the 3-point structure constant in \eqref{definition-normalization}. Such structure constant indeed reproduces the different $\cosh$ and $\sinh$ dependences in each regime, signaling this is a genuine difference between the two. On the other hand, we could easily cook up a normalization $N(\alpha)$ such that the timelike structure constant $U(\alpha)\sim\cosh(\pi s(2\alpha-q))$. However, such a choice would imply that the operators are normalized with an $s$-dependent function, and therefore that also the bulk correlators become $s$-dependent.
In other words, while one could choose a normalization for which both space and timelike 1-point structure constants are the analytic continuation of each other, such a choice entails a drastic change of the bulk theory. 

Another important difference between the spacelike and the timelike 1-point functions is the $s$ or $\mu_B$ dependence. Concretely, the only $s$ dependence of the FZZT solution \eqref{FZZT} is in the argument of the $\cosh [\pi s(2a-Q)]$, while our timelike solution \eqref{1-pnt-fn-normalised} has an $s$ dependence as $\sinh[\pi s(2\a-q)]/ s$. While this may seem bizarre at first sight, it responds to the expectation that the 1-point structure constant should be invariant under $s\rightarrow -s$. Indeed, the equations relating $s$ to $\mu_B$ in both regimes are invariant under such transformation of $s$. There is hence no physical meaning to its sign and is just to be expected the 1-point structure constants be invariant under this transformation as well. The different dependence on $s$ in each regime is hence a consequence again of the relative minus sign between the timelike and spacelike reflection coefficients, which in the spacelike case determines an even function of $s$, $\cosh(s\dots)$, while in the timelike case determines an odd function of $s$, $\sinh(s\dots)$. (The question of the limit $\mu \rightarrow 0$ or $s\rightarrow \infty$ is briefly addressed below.)

 Having said this, it is also worth noting that the $s$-factor in the denominator of our timelike solution is subject to our choice of normalization \eqref{normalization-1pnt-fn-T} (indeed such a factor does not appear in the non-normalized structure constant \eqref{unnormalised-1-pnt-fn-timelike}). It would be interesting to explore different normalizations where no such additional $s$-dependence appears.\footnote{One particularity of our normalization is that, while the 1-point structure constant is fixed to satisfy the perturbative answer for the case where $\alpha\rightarrow q/2$ or $n=0$, as imposed by \eqref{normalization-1pnt-fn} and \eqref{normalization-1pnt-fn-T}, the resulting expression \eqref{1-pnt-fn-normalised} exhibits a pole for all other $\alpha=q/2+n\beta$ with $n=1,2,3,...$, due to the factor $\Gamma(1+\beta^{-1}(q-2\a))=\Gamma(1-n)$, and the residue does not coincide with the perturbative answer for these cases, given by \eqref{Liouv-corr-Cgas-corr-boundary}. This fact may also be a good reason to search for other normalizations.}

We can further compare the two equations for the $s$ parameter in each regime  \eqref{eq:s_muB_SL} and \eqref{eq:s_muB_TL}:
\begin{equation}\label{label}
	\cosh (\pi b s) = \mu_B \,\sqrt{\frac{\sin(\pi b^2)}{\mu}} \qquad \cosh (\pi \b s) = -\mu_B\,\sqrt{\frac{\sin(\pi \b^2)}{-\mu}}.
\end{equation}
Following the discussion at the end of section \ref{sec:1-pnt-fn}, for generic value of $b$, the spacelike boundary solution requires $\mu>0$, while the timelike one requires  $\mu<0$ (assuming that both bulk and boundary cosmological constants are real).
In the spacelike semiclassical limit $b\rightarrow 0$, bulk and boundary cosmological constants scale as $\mu,\mu_B\sim b^{-2}$, as follows from the equation of motion and boundary condition for the classical field $\phi_{c}=2b\phi$, 
\be
\partial\bar\partial \phi_c=2\pi\mu b^2e^{\phi_c},\qquad \qquad i(\partial-\bar\partial)\phi_c=4\pi\mu_B b^2 e^{\phi_c/2}.
\ee
The semiclassical limit of the above spacelike equation is then compatible with $\mu>0$. 
In the timelike regime, the bulk cosmological constant scales instead as $\mu\sim -\b^{-2}$, since the equation of motion for the classical field $\chi_c=2\b\chi$ has now an additional minus sign in front  of the derivative term:
\be
-\partial\bar\partial \chi_c=2\pi\mu \b^2e^{\chi_c}.
\ee
 This minus sign compensates for the one in the timelike equation for $s$ in \eqref{label}.

Finally, we may compare the $\a$-independent normalizations between the two solutions. While our timelike solution \eqref{1-pnt-fn-normalised} is finite in the limit $\a\rightarrow q/2$, the FZZT solution \eqref{FZZT} is divergent at $a=Q/2$. In this case then, the normalization needs to be defined with the residue. In particular, FZZT demand
\be
\underset{a=Q/2}{\text{Res}}\,U_{\small\sub{\text{FZZT}}}(a)= 1,
\ee
instead of our limit condition $\lim_{\a\rightarrow q/2}U(\a)= \frac{1}{2\beta}$ in the timelike regime (the $1/2\b$ instead of the $1$ factor is just due to a different normalistion of the free correlators and path integral compared to ours).

It is interesting to note that in the spacelike regime $\mathcal R_{\sub{\text{DOZZ}}}(Q/2)=-1$, which implies that the primary $V_{Q/2}$ exactly vanishes. So it is in fact somewhat surprising that the FZZT solution does not vanish, but instead diverges, for this charge value. In the timelike regime instead, $\mathcal R(q/2)=1$ and the primary $V_{q/2}$ is finite, consistent with our 1-point function solution having a finite $\a\rightarrow q/2$ limit.

\vspace{5mm}

We now compare our solution for the timelike boundary 2-point structure constant  to the spacelike FZZT solution \cite{Fateev:2000ik,Teschner:2000md}. The latter reads
\begin{equation}\label{eq:bdry2pt_SL}
	\begin{aligned}
		D_{\small\sub{\text{FZZT}}}(d|s_1,s_2) = \left[\pi\mu\gamma(b^2)b^{2-2b^2}\right]^{(Q-2d)/2b}\frac{\Gamma_b(2d-Q)}{\Gamma_b(Q-2d)}\,\prod_{\pm\pm'}\frac{1}{S_b(d\pm i\tfrac{s_1\pm's_2}{2})}.
	\end{aligned}
\end{equation}
where $d$ is the charge of the boundary operators in the spacelike boundary 2-point function, $\langle B_{d_1}^{\mu_1\mu_2}(x)B_{d_2}^{\mu_2\mu_1}(0)\rangle$.  This expression was also obtained as a solution to shift equations analogous to those used in section \ref{sec:bdry-2pnt-fn} but valid in the spacelike regime. The first thing to notice is that the analytic continuation of this solution to the timelike regime is not defined because $\Gamma_b(x)$ has simple poles for $x = -mb-nb^{-1}$ where $m$ and $n$ are non-negative integers, so that when $b$ is taken to be purely imaginary, infinitely many poles accumulate for certain imaginary values of $x$.

Comparing with our solution for $D(\d|s_1,s_2)$ \eqref{bdry2pt_TL}, we note that the two structure constants are not the analytic continuation of each other: the arguments in the double-sine functions in the two expressions are shifted by $\b$ terms. This is also the case for the two factors of double Gamma functions. This is reminiscent of what happens for the bulk 3-point structure constants: the arguments of the $\Upsilon_\b$  functions in the timelike expression \eqref{TL-structure-cntt} exhibit $\b$-shifts with respect to those in the DOZZ formula. 

The spacelike expression \eqref{eq:bdry2pt_SL} is such that $D_{\small\sub{\text{FZZT}}}(Q/2|s_1,s_2)=-1$, while our timelike expression satisfies $D(q/2|s_1,s_2)=1$. This behaviour is analogous to what happens with the bulk reflection coefficients as explained above,  $\mathcal R_{\sub{\text{DOZZ}}}(Q/2)=-1$ but $\mathcal R(q/2)=1$, confirming the good role of $D$ as boundary reflection coefficient.

\vspace{5mm}

It is also worth comparing our results to related ones obtained earlier in the literature. In \cite{Gutperle:2003xf}, Gutperle and Strominger obtained the bulk 1-point and boundary 2-point structure constants of the timelike Liouville theory but with the bulk cosmological constant turned off $\mu=0$, and by analytic continuation of the spacelike expressions of such theory. Since $\mu=0$,  their 1-point structure constant exhibits a power law dependence on $\mu_B$ consistent with the relation \eqref{eq:s_muB_TL} and the dependence on the operator charge is an inverse $\sin$ function. The analytic continuation of the spacelike boundary 2-point function was found to diverge, and an integration contour was prescribed to circumvent the divergence. If we want to compare our results to theirs, we need to take the limit $\mu\rightarrow 0$, or equivalently $s\rightarrow \infty$, of our expressions. However, such limit of our 1-point solution \eqref{1-pnt-fn-normalised} goes to zero due to the factor of $s$ in the denominator. In order to to reproduce the results in \cite{Gutperle:2003xf} we would need to pick a normalization factor that does not depend $s$, i.e., the proportionality constant $C$ in \eqref{unnormalised-1-pnt-fn-timelike} is $s$-independent (though it may still depend on $\b$).

\subsection*{Outlook}
It would be very interesting to find the timelike analog of the ZZ boundary conditions. Spacelike ZZ boundary conditions are parametrized by a pair of positive integers $m,n$, so that for instance, the disk one-point structure constant \cite{Zamolodchikov:2001ah} in this boundary condition is given by
\be
U_{\sub{\text{ZZ}}}^{(m,n)}(a)=\frac{\sin(\pi b^{-1}Q)\,\sin(\pi \,m\,b^{-1}(2a-Q))}{\sin(\pi b^{-1}(2a-Q))\,\sin(\pi \,m\, b^{-1}Q)}\,\frac{\sin(\pi b\,Q)\,\sin(\pi \,n\,b(2a-Q))}{\sin(\pi b (2a-Q))\,\sin(\pi \,n\, bQ)}\,U_{\sub{\text{ZZ}}}^{(1,1)}(a),
\ee
where
\begin{equation}
	U_{\sub{\text{ZZ}}}^{(1,1)}(a) = \frac{\left[\pi \mu \gamma(b^2)\right]^{-a/b}\Gamma(1+b^2)\Gamma(1+b^{-2})Q}{(Q-2a)\Gamma(b(Q-2a))\Gamma(b^{-1}(Q-2a))}
\end{equation}
While the analytic continuation of this solution is well-defined just as the FZZT solution, one can verify that the analytic continuation does not satisfy the required reflection property. 

It would also be very interesting to  explore the geometrical interpretation of such a solution. The ZZ 1-point function entails setting Dirichlet boundary conditions for the Liouville field at infinity, and hence corresponds to a D0-brane localised in the Liouville field spacelike direction. An analogous timelike solution could correspond to a brane localised in the field timelike direction. 
It would also be interesting to check if such a possible ZZ-like solution is related to our result for the 1-point function \eqref{1-pnt-fn-normalised}, just as the spacelike FZZT and ZZ solutions are \cite{Martinec:2003ka,McGreevy:2003ep}. More importantly, with these two solutions in hand, we should study the spectrum of boundary states of this theory in detail. Together with the search of ZZ-like solutions, this is a necessary next step to take.

In \cite{McGreevy:2003kb}, a connection was established between $1+1$-dimensional string theory and the $c=1$  matrix quantum mechanics. In particular, it was proposed  that this matrix model corresponds to the theory of unstable $D0$ branes in the minimal string theory. This proposal was  based on the quantitative match between  the rate of closed string emission produced by a rolling eigenvalue of the matrix quantum mechanics, and that produced by a rolling tachyon in the string theory, where the latter is computed with the analytic continuation of the FZZT 1-point structure constant. It would be very interesting to explore similar avenues with our timelike 1-point function solution, and eventually see if they can shed any light on a possible microscopic description of timelike Liouville gravity.

As for boundary timelike Liouville theory as a BCFT, in order to have a complete description we would need to also compute the bulk-boundary and the boundary 3-point structure constants. As for the latter, it would be interesting to check its relation to the boundary 2-point structure constant we have computed \eqref{bdry2pt_TL}. Just as it happens with their bulk analogs, we would not expect the limit of the boundary 3-point structure constant when one of the charges is taken to vanish, to give a diagonal expression (though we would expect it to be related the the boundary 2-point constant when evaluated on two equal charges just as \eqref{reflection-coef-3pntfn} for the bulk). In other words, we would not expect the boundary operator with vanishing dimension to be the identity, but rather a non-degenerate operator. 
More generally, given that timelike Liouville theory, just as its spacelike counterpart, lends itself easily to exact solutions, any explorations of its timelike boundary description has the potential to teach us a lot about non-unitary BCFTs.

\section*{Acknowledgements}

We would like to thank Atish Dabholkar, Harold Erbin, and Piotr Su\l{}kowski. We are specially thankful to Sameer Murthy and Sylvain Ribault, for clarifications on several points and comments on the draft. A.B. would like to thank the International Centre for Theoretical Physics, Trieste for its hospitality during the first stages of this project. The work of  T.B. is supported by STFC grants ST/P000258/1 and ST/T000759/1. The work of A.B. is supported by the TEAM programme of the Foundation for Polish Science co-financed by the European Union under the European Regional Development
Fund (POIR.04.04.00-00-5C55/17-00).

\appendix

\section{Special functions}\label{app:special functions}

In this appendix, we list the special functions relevant for our results and calculations.

\subsubsection*{The little gamma function ($\gamma$)}
\begin{equation}
\gamma(x):= \frac{\Gamma(x)}{\Gamma(1-x)}.
\end{equation}
\begin{equation*}
\gamma(x)\gamma(1-x)=1,\qquad 
\gamma(x)\gamma(-x)=-x^{-2}
\end{equation*}
\subsubsection*{The double Gamma function ($\Gamma_\b$)}
The double Gamma function $\Gamma_\b \;(= \Gamma_{\b^{-1}})$ can be defined by the following integral representation, valid for $\mathrm{Re}\,x>0$
\small\begin{equation}\label{eq:Gamma_b_def}
\log \Gamma_\b(x) = \int_{0}^{\infty}\frac{dt}{t}\left[\frac{e^{-xt}-e^{-(\b^{-1}+\b)t/2}}{(1-e^{-\b t})(1-e^{-\b^{-1}t})} - \frac{1}{2}\left(\tfrac{\b^{-1}+\b}{2}-x\right)^2e^{-t}-\frac{1}{t}\left(\tfrac{\b^{-1}+\b}{2}-x\right)\right]
\end{equation}\normalsize
and by analytic continuation elsewhere on the complex plane. This function is meromorphic with simple poles at $x = -m\b -n\b^{-1}$ where $m$ and $n$ are non-negative integers, whereas $1/\Gamma_b(x)$ is an entire function. It follows from the definition that $\Gamma_\b(\frac{\b^{-1}+\b}{2})=1$.

Crucial to our calculations are the following shift relations:
\begin{equation}\label{eq:Gamma_b_shift}
\begin{aligned}
\Gamma_\b(x +\b)= \frac{\sqrt{2\pi}\b^{x\b-\tfrac{1}{2}}}{\Gamma(\b x)}\Gamma_\b(x),\qquad
\Gamma_\b(x +\b^{-1})= \frac{\sqrt{2\pi}\b^{-x\b^{-1}+\tfrac{1}{2}}}{\Gamma(\b^{-1}x)}\Gamma_\b(x).
\end{aligned}
\end{equation}
Further details and proofs of these claims can be found in Appendix A of \cite{Nakayama:2004vk} and Appendix A of \cite{Jimbo:1996ss}.

\subsubsection*{The Upsilon function ($\Ups_\b$)}

This can be defined in terms of the double Gamma function as
\begin{equation}\label{eq:Ups_def}
\Ups_\b(x) := \frac{1}{\Gamma_\b(x)\Gamma_\b(\b^{-1}+\b-x)}.
\end{equation}
Clearly, $\Ups_\b(x) = \Ups_\b(\b^{-1}+\b-x)$. 


The shift formulae for the Upsilon function follow from \eqref{eq:Gamma_b_shift}:
\begin{equation}\label{eq:UpsShifts}
\begin{aligned}
\Ups_\b(x+\b) = \gamma(\b x)\,\b^{1-2\b x}\,\Ups_\b(x),\qquad
\Ups_\b(x+\b^{-1}) = \gamma(\b^{-1}x)\,\b^{2\b^{-1}x-1}\,\Ups_\b(x)
\end{aligned}
\end{equation}
The following is often useful:
\begin{equation*}
\begin{aligned}
\Ups_\b(-x)=\Ups_\b(x+\b+\b^{-1}) &= -x^2\,\gamma(\b x)\,\gamma(\b^{-1}x)\,\b^{2(\b^{-1}-\b)x}\,\Ups_\b(x)\\
&= -\frac{\Gamma(\b x)\,\Gamma(\b^{-1}x)}{\Gamma(-\b x)\Gamma(-\b^{-1}x)} \b^{2(\b^{-1}-\b)x}\,\Ups_\b(x).
\end{aligned}
\end{equation*}

It follows from the definition \eqref{eq:Ups_def} that $\Ups_\b(x)$ is an entire function with simple zeros at $x = -m\b -n\b^{-1}$ and $x = (m+1)\b +(n+1)\b^{-1}$, where $m$ and $n$ are non-negative integers, and that $\Ups_\b(\frac{\b^{-1}+\beta}{2})=1$.

Other useful shift relations are:
\begin{align}
\Upsilon_\beta ((\b^{-1}-\b)-x)=\Upsilon_\beta(x+2\beta),\qquad \Upsilon_\beta (x-(\b^{-1}-\b))=\Upsilon_\beta(\frac{2}{\beta}-x).
\end{align}

	\subsubsection*{The double-sine function ($S_\b$)}
While the product of $\Gamma_\b(x)$ and $\Gamma_\b(\b^{-1}+\b-x)$ gives the Upsilon function, their ratio defines the double sine function:
\begin{equation}\label{double-sine-fn}
S_\b(x) := \frac{\Gamma_\b(x)}{\Gamma_\b(\b^{-1}+\b-x)}.
\end{equation}
Its name is justified by the following shift relations:
\begin{equation}\label{eq:double_sin_shifts}
\begin{aligned}
S_\b(x+\b) = 2\sin(\pi \b x)\,S_\b(x),\qquad
S_\b(x+\b^{-1})&= 2\sin(\pi \b^{-1} x)\,S_\b(x).
\end{aligned}
\end{equation}

\subsubsection*{The Hypergeometric function ($_2F_1$)}
The four-point conformal block on a sphere with a level two degenerate operator satisfies a differential equation which can be brought into the form of the following hypergeometric equation:
\begin{equation}\label{eq:hypergeometric}
x(1-x)\frac{d^2f}{dx^2} + \left[C-(A+B+1)x\right]\frac{df}{dx}-ABf = 0.
\end{equation}
For generic values of $A,B,C$, its two linearly independent solutions, expanded about $x=0$, are
\begin{equation*}
\begin{aligned}
_2F_1(A,B;C;x),\qquad \qquad\qquad(1-x)^{1-C}\,_2F_1(1+A-C,1+B-C;2-C;x),
\end{aligned}
\end{equation*}
where  $_2F_1(A,B;C;x) = \,_2F_1(B,A;C;x)$ is the hypergeometric function.

We note a couple of identities that we use: the Euler transformation
\begin{equation}\label{eq:hypereuler}
_2F_1(A,B;C;x) = (1-x)^{C-A-B}\,_2F_1(C-A,C-B;C;x) 
\end{equation}
and the following connection formula that relates a $_2F_1$  expanded around $x=0$ to a linear combination of a pair of $_2F_1$s expanded around $x=1$:
\begin{equation}\label{eq:hypertrans}
\begin{aligned}
_2F_1(A,B;&C;x)=\frac{\Gamma(C)\Gamma(C-A-B)}{\Gamma(C-A)\Gamma(C-B)}\,_2F_1(A,B;A+B+1-C;1-x)\\ &+\frac{\Gamma(C)\Gamma(A+B-C)}{\Gamma(A)\Gamma(B)}(1-x)^{C-A-B}\,_2F_1(C-A,C-B;1+C-A-B;1-x)
\end{aligned}
\end{equation}

	\section{Degenerate conformal blocks}\label{app:conformal blocks}
	
In this appendix we derive the 4-point degenerate conformal blocks required for the computation of various shift equations that appear in this work.

Consider a conformal 4-point correlator of primary operators with an insertion of either one of the two level-2 degenerate operators, 
\begin{align}\label{degenerate-stts}
&V_{\langle 1,2\rangle}\,:\,\, \alpha_{1,2}=\frac{\beta}{2},\quad \Delta_{1,2}=-\frac{1}{2}+\frac{3}{4}\beta^2,\qquad
&V_{\langle 2,1\rangle}\,:\,\,\alpha_{2,1}=-\frac{1}{2\beta},\quad \Delta_{2,1}=-\frac{1}{2}+\frac{3}{4\beta^2}.
\end{align}
We focus for now on the first degenerate. Consider in particular the insertions at specific points:
\be
\F(z) = \langle V_{\a_3}(\infty)V_{\a_2}(1)V_{\langle1,2\rangle}(z,\bar{z})V_{\a_1}(0)\rangle.
\ee
 This correlator satisfies the following BPZ equation:
\small\begin{equation}\label{eq:4pt_diffeqn}
\left[\frac{1}{\b^2}\frac{d^2}{dz^2}+\left(\frac{1}{z-1}+\frac{1}{z}\right)\frac{d}{dz} - \frac{\Delta_{2}}{(z-1)^2}- \frac{\Delta_{1}}{z^2}+\frac{\Delta_{1,2}+\Delta_{1}+\Delta_2-\Delta_3}{z(z-1)}\right]\F(z)=0
\end{equation}\normalsize
along with its anti-holomorphic counterpart. After some redefinitions, this equation can be recast into the form of a hypergeometric equation \eqref{eq:hypergeometric}, its solutions are given in terms of the hypergeometric function $_2F_1$, and are called degenerate conformal blocks.

The solutions to this equation have singularities at either $z=0$, $z=1$, or $z=\infty$. Singularities of conformal correlators correspond to operators coming close to each other, and so each of these three singularities corresponds to each of the three s, t, and u-channels. For each singularity point, there are two solutions, corresponding to the two possible states being exchanged in that channel.

The s-channel degenerate conformal blocks correspond to solutions singular at $z\rightarrow0$, 
\begin{equation}\label{conf-bloc-sch-timelike}
\begin{aligned}
\F^{s}_+(z)&\equiv\mathcal{F}\left[\begin{smallmatrix}
\b/2& \a_2\\
\a_1& \a_3
\end{smallmatrix};{\scriptstyle{\a_1+}}\tfrac{\b}{2};{\scriptstyle{z}}\right]=z^{\a_1\b}(1-z)^{\a_2\b}\,_2F_1(A,B;C;z),\\
\F^{s}_-(z)&\equiv \mathcal{F}\left[\begin{smallmatrix}
\b/2& \a_2\\
\a_1& \a_3
\end{smallmatrix};{\scriptstyle{\a_1-}}\tfrac{\b}{2};{\scriptstyle{z}}\right]=z^{1-\a_1\b-\b^2}(1-z)^{1-\a_2\b-\b^2}\,_2F_1(1-A,1-B;2-C;z),
\end{aligned}
\end{equation}
and the t-channel blocks correspond to solutions singular at $z\rightarrow 1$, 
\begin{equation}\label{conf-bloc-tch-timelike}
\begin{aligned}
\F^t_+(z)&\equiv\mathcal{F}\left[\begin{smallmatrix}
\b/2& \a_1\\
\a_2& \a_3
\end{smallmatrix};\scriptstyle{\a_2+}\tfrac{\b}{2};1-z\right] =z^{\a_1\b}(1-z)^{\a_2\b}\,_2F_1(A,B;1+A+B-C;1-z),\\
\F^t_-(z)&\equiv\mathcal{F}\left[\begin{smallmatrix}
\b/2& \a_1\\
\a_2& \a_3
\end{smallmatrix};\scriptstyle{\a_2-}\tfrac{\b}{2};1-z\right] =z^{1-\a_1\b-\b^2}(1-z)^{1-\a_2\b-\b^2}\,_2F_1(1-A,1-B;1+C-A-B;1-z),
\end{aligned}
\end{equation}
where
\begin{equation}\label{ABC}
\begin{aligned}
A=-1+(\a_1+\a_2+\a_3)\b+\frac{3\b^2}{2},\qquad
B=(\a_1+\a_2-\a_3)\b+\frac{\b^2}{2},\qquad
C=2\a_1\b+\b^2.
\end{aligned}
\end{equation}
One can verify that each of $\F^s_\pm$ and $\F^t_\pm$ are solutions to the differential equation \eqref{eq:4pt_diffeqn}. One can also verify that the parameters in $\F^s_\pm$ and $\F^t_\pm$ are identical up to the exchange of $\a_1$ and $\a_2$, and $z\rightarrow 1-z$. As a further check, note that the above expressions satisfy the property (see for example\cite{2014arXiv1406.4290R})
\begin{equation*}
\mathcal{F}\left[\begin{smallmatrix}
\a_2& \a_3\\
\a_1& \a_4
\end{smallmatrix};{\scriptstyle{\a}};\scriptstyle{z}\right]\stackrel{z\rightarrow 0}{=}z^{\Delta_\a-\Delta_{1} - \Delta_{2}}(1 + O(z)).
\end{equation*}

Since the hypergeometric equation has only two independent solutions, the two solutions in each channel are related to those in either one of the other two channels. Using identities \eqref{eq:hypereuler} and \eqref{eq:hypertrans} we see that
\begin{equation}\label{eq:blockconn}
\F^s_i(z) = \sum_{j=\pm}\mathcal{B}_{ij}\, \F^t_j(z), \qquad i=\pm,
\end{equation} 
where
\begin{equation}\label{fusing-matrix-elements}
\begin{aligned}
\mathcal{B}_{++}&= \frac{\Gamma(C)\Gamma(C-A-B)}{\Gamma(C-A)\Gamma(C-B)} , &&\mathcal{B}_{+-}=\frac{\Gamma(C)\Gamma(A+B-C)}{\Gamma(A)\Gamma(B)},\\
\mathcal{B}_{-+}&= \frac{\Gamma(2-C)\Gamma(C-A-B)}{\Gamma(1-A)\Gamma(1-B)}, &&\mathcal{B}_{--}=\frac{\Gamma(2-C)\Gamma(A+B-C)}{\Gamma(1+A-C)\Gamma(1+B-C)}.
\end{aligned}
\end{equation}
The matrix $\mathcal B_{ij}$ is called the degenerate fusing matrix.

The above conformal blocks and their fusing matrix can be used to derive shift equations for the different CFT correlators data. Since they only encode the kinematical part of the corresponding 4-point functions, they need to be multiplied with the corresponding dynamical data, which upon channel-decomposition is given in terms of structure constants and OPE coefficients. Because the 4-point functions considered contain a degenerate operator, the channel decomposition involves only two intermediate states. This can be for instance used \cite{Teschner:1995yf} to obtain the 3-point function structure constants.
Further, in the case one of the three generic insertions is also given by the same level-2 degenerate, the resulting $\mathcal B_{ij}$ are required to derive shift equations for the bulk 1-point function, as done in section \ref{sec:1-pnt-fn}.

\section{Coulomb gas approach}\label{app:DF-integrals}

In this section we summarize a method used for computing certain special correlators in Liouville theory, referred to as the Coulomb gas approach, the method of screening integrals or Dotsenko-Fateev integrals, or the perturbative approach. References include \cite{Dotsenko:1984nm,Fateev:2000ik,Zamolodchikov:1995aa}. Our choice of normalization for the correlation functions is based on this method.

This method identifies a relation between Liouville correlators and Coulomb gas correlators on the sphere for certain combinations of the charges of the insertion operators. Such a relation follows from integrating the zero mode in the path integral expression of the correlator. Indeed, if we separate the Liouville field as $\chi=\bar\chi+\chi_{\0}$, where $\chi_{\0}$ is the zero mode, then a generic correlator
\be\label{correlator-path-integral}
\langle \prod\limits_{i=1}^n V_{\alpha_i}(z_i)\rangle=\int \mathcal D {\bar\chi}\,\mathcal D\chi_{\0}\,e^{-S_{tL}[\bar\chi+\chi_{\0}]}\,e^{-2\alpha_t\,\chi_{\0}}\,\prod\limits_{i=1}^n e^{-2\alpha_i\,\bar\chi(z_i)},
\ee
where $\a_t=\sum{\a_i}$ is the total Liouville charge.
Integrating $\chi_{\0}$ using the integral expression for the Gamma function, one obtains
\begin{align}\label{Liouville-correls-Cg-correls}
	\langle \prod\limits_{i} V_{\alpha_i}(z_i)\rangle
	&=\frac{\Gamma(-n)}{2\beta}\,\mu^n\,\langle \prod\limits_{i} e^{-2\alpha_i\bar\chi(z_i)}\,\left(\int d^2w\, e^{2\beta\bar\chi(w)}\right)^n\rangle_{\scriptscriptstyle{0}},
\end{align}
where
\be
n=\frac{\a_t-q}{\b},
\ee
and  the subscript on the right-hand-side bracket indicates the correlator is evaluated on the Coulomb gas theory. The integral of the zero mode effectively brings the exponential interaction term from the original action down to the path integrand, turning the action into that of a Coulomb gas and the interaction term into correlator insertions.

Though this relation is derived on the sphere, it holds in the complex plane as well since the relation between the two is only a conformal transformation. The correlator on the right-hand side of \eqref{Liouville-correls-Cg-correls} is then evaluated on a (timelike) free scalar theory, its generic expression is
\be\label{free-correl-timelike}
\langle \prod_i V_{\alpha_i}(z_i)\rangle_{\scriptscriptstyle{0}}=\prod_{i>j}\,|z_{ij}|^{4\alpha_i\,\alpha_j}.
\ee
Notice that the power of the dimensions is positive since this correlator is 
for a timelike field, whose free propagator  $\sim\log|z_{ij}|^2$.

To be able to use the expression for the free correlator in the above relation \eqref{Liouville-correls-Cg-correls}, $n$ must be a non-negative integer, but in that case $\Gamma(-n)$ diverges. However, the perturbative expansion of the path integral \eqref{correlator-path-integral} in $\mu$, which would give \eqref{Liouville-correls-Cg-correls} at order $n$, suggests to rewrite the above as\footnote{In the spacelike Liouville regime, the DOZZ formula for the 3-point structure constant is divergent when one of the operators is a degenerate one, namely when $a_{\langle r,s\rangle}=Q/2-(r/b +sb)/2$, which is the kind of situation where $n=(Q-a_t)/b$ is a non-negative integer. In such cases, the 3-point structure constant is obtained from the residue of the corresponding DOZZ formula, which further justifies substituting $\Gamma(-n)$ by its residue $-(-1)^n/n!$.}
	\begin{align}\label{Liouville-correls-Cg-correls-residue}
		\langle \prod_i V_{\alpha_i}(z_i)\rangle
		&=-\frac{1}{2\beta}\,\frac{(-\mu)^n}{n!}\,\langle \prod_i e^{-2\alpha_i\bar\chi(z_i)}\,\left(\int d^2w\, e^{2\beta\bar\chi(w)}\right)^n\rangle_{\scriptscriptstyle{0}},
	\end{align}
	namely it suggests to replace $\Gamma(-n)$ by its residue $-\frac{(-1)^n}{n!}$.
This expression is hence typically referred to as the perturbative expression, and the integrated operators are called screening operators, since they ensure that the total charge of the Coulomb gas correlator is equal to $q$. This establishes our choice of normalization: it is such that the above expression \eqref{Liouville-correls-Cg-correls-residue} with the free correlators given by \eqref{free-correl-timelike} is satisfied for Liouville correlators with $\a_t=q+n\b$.

This expression can be for instance used to determine the OPE coefficients $C_\pm(\a)$, in $V_\a\,V_{\langle1,2\rangle}\sim C_\pm(\a) V_{\a\pm \b/2}$, used in section \ref{sec:timelike boundary Liouville} and in appendix \ref{app-bulk-correls}. These two coefficients are given by 3-point structure constants as usual:
{\small\begin{align}\label{OPE-coefs-3pntfn}
		C_+(\alpha)=\frac{   C( \alpha,\frac{\beta}{2},\alpha+\frac{\beta}{2})}{   G(\alpha+\frac{\beta}{2})}=C(\alpha, \frac{\beta}{2},q-\alpha-\frac{\beta}{2}),\qquad\quad
		C_-(\alpha)=\frac{   C( \alpha,\frac{\beta}{2},\alpha-\frac{\beta}{2})}{   G(\alpha-\frac{\beta}{2})}=C( \alpha,\frac{\beta}{2},q-\alpha+\frac{\beta}{2}),
		\end{align}}
where on the second step we have used the relation between the bulk 2-point function and the reflection coefficient $G(\a)=\R(\a)$ \eqref{2pnt-fn-timelike}. 

For $C_+(\a)$, the charges of the corresponding 3-point function are such that $\a_t=q$, hence $n=0$ and the 3-point function requires no insertion of screening operators in \eqref{Liouville-correls-Cg-correls-residue}. As a consequence, it is independent of $\alpha$, and with the normalization chosen for the free-field correlators, 
\be\label{OPE-coef_C+}
C_+(\a)=-\frac{1}{2\b}.
\ee

In the case of $C_-(\a)$, the charges of the corresponding 3-point function satisfy $n=1$, and hence requires the insertion of one screening operator:
\begin{equation}\label{C-OPE-coef-screening}
C_-(\a)= \frac{\mu}{2\beta}\int d^2w\,\left\langle V_{\a}(0)V_{\frac{\b}{2}}(1)V_{q-\a+\frac{\b}{2}}(\infty)V_{-\b}(w) \right\rangle_{\scriptscriptstyle{0}}= \frac{\mu}{2\b}\int d^2w\, |1-w|^{-2\b^2}|w|^{-4 \a\b}.
\end{equation}
The resulting integral  can be performed using the Dotsenko-Fateev integrals formulae \cite{Dotsenko:1985hi,Dotsenko:1984ad}, 
\begin{equation*}
\int d^2w\,|w|^{2(m-1)}|1-w|^{2(l-1)} = \frac{\pi\gamma(m)\gamma(l)}{\gamma(m+l)}.
\end{equation*}
The OPE coefficient becomes
\be\label{residue-condition-TL-3pntfn}
C_-(\a)=\frac{\pi\mu}{2\b\,\gamma(\b^2)}\frac{\gamma(2\a\b+\b^2-1)}{\gamma(2\a\b)}.
\ee
Notice that the above choice of normalization is equivalent to fixing the normalization of the 3-point function structure constant to satisfy,
\be\label{residue-condition-TL-3pnt-fn}
C(\alpha_1,\alpha_2,\alpha_3)\underset{\alpha_t=q}{=} -\frac{1}{2\beta},
\ee
and more generically
\be
C(\a_1,\a_2,\a_3)\underset{\alpha_t=q+n\b}{=}\,-\frac{1}{2\b}\frac{(-\mu)^n}{n!}\int d^2w_1...d^2w_n \,\langle V_{\a_1}(0)V_{\a_2}(1)V_{\a_3}(\infty)\prod_i^n V_{-\b}(w_i)\rangle_{\scriptscriptstyle{0}}.
\ee
The above choice of normalization, namely the factor of $-1/2\beta$ in \eqref{OPE-coef_C+}, fixes then the $\a_i$-independent normalization of
the 3-point structure constant. In particular, notice that the above expressions for the OPE coefficients, \eqref{OPE-coef_C+} and \eqref{residue-condition-TL-3pntfn}, exactly follow from \eqref{OPE-coefs-3pntfn} with the expression for 3-point structure constant \eqref{TL-structure-cntt}. This is in sharp contrast to the spacelike regime, where the 3-point structure constant as given by the DOZZ formula diverges for the combination of charges required for the analogous OPE coefficients. There then, the perturbative method is specially useful.


The above relation between Liouville and free-theory correlators for special cases of the Liouville charges can also be derived in the case where the theory is placed on the upper half plane, in the presence of a boundary cosmological constant $\mu_B$, and boundary operators  on the boundary along the real axis. In this case, 
\begin{align}\label{Liouv-corr-Cgas-corr-boundary}
 \langle\prod_i V_{\a_i}(z_i)\,\prod_j B_{\d_j}&(x_j)\rangle= \frac{1}{2\b}\,\sum_{k=0}^{\lfloor n/2 \rfloor}\frac{(-\mu)^k(-\mu_B)^{n-2k}}{k!(n-2k)!} \\
 &
 \times\,\Big\langle \prod_i V_{\a_i}(z_i)\,\prod_j B_{\d_j}(x_j)\left(\int_{\sub\text{Im}\,z>0}d^2w_i\,V_{-\b}(w_i)\right)^k\left( \int_{-\infty}^\infty dy_j\,B_{-\b}(y_j)\right)^{n-2k} \Big\rangle_{\sub 0},\nn\end{align}
 where now
\be
n=\frac{2\a_t+\d_t-q}{\b}.
\ee
Again, the above expression is only valid when $n$ is a non-negative integer. Just as \eqref{Liouville-correls-Cg-correls-residue}, this expression can be thought of as perturbative in the sense that it can be interpreted as following from a perturbative expansion in $\mu$, $\mu_B$ up to a total power of $n$. This inspires the substitution of
 $\Gamma(2k-n)$ by $\frac{(-1)^{n-2k}}{2(n-2k)!}$, i.e. the divergent Gamma function by its residue. The integrated correlator on the right-hand side is evaluated on the free theory on the upper half plane, which reads
	\begin{equation}\label{free-correl-bdry}
\begin{aligned}
\langle \prod_i V_{\a_i}(z_i)\,\prod_j B_{\d_j}(x_j)\rangle_{\sub 0}= \frac{\left(\prod\limits_{i}|z_i-\bar{z}_i|^{2\a_i^2}\right)\left(\prod\limits_{i,j}|z_i-x_j|^{4\a_i \d_j}\right)}{\left(\prod\limits_{i>j}|x_i-x_j|^{-2\d_i \d_j}\right)\left(\prod\limits_{i>j}|(z_i-z_j)(z_i-\bar{z}_j)|^{-4\a_i \a_j}\right)}.
\end{aligned}
\end{equation}
This expression is used in section \ref{sec:1-pnt-fn} to determine the normalization of the bulk 1-point structure constant.

\section{Timelike bulk correlators}\label{app-bulk-correls}

In this appendix we review the derivation of the 2- and 3-point functions and the normalizations chosen. We closely follow \cite{Ribault:2014hia} in that we try to distinguish between the parts of the structure constants that are required by the bootstrap, and those that are fixed by choosing the normalization.

The general form of the 2-point function is
\be
\langle V_{\alpha_1}(z_1)\,V_{\alpha_2}(z_2)\rangle= 2\pi\frac{   G(\alpha_1)\,\left[\delta(\alpha_1-\alpha_2)+   \R(\alpha_2)\delta(q-\alpha_1-\alpha_2)\right]}{|z_{12}|^{2(\Delta_1+\Delta_2)}},
\ee
with the reflection coefficient and the 2-point structure constant to be determined.

If we impose $\langle V_{\alpha_1}\,V_{\alpha_2}\rangle=   R(\alpha_1)\,\langle V_{q-\alpha_1}\,V_{\alpha_2}\rangle$, we find
\be
\frac{   G(\alpha)}{   G(q-\alpha)}=\frac{   \R(\alpha)}{   \R(q-\alpha)},
\ee
which suggests the identification of the 2-point function and the reflection coefficient up to an $\alpha$-independent constant. In the case of spacelike Liouville, this constant is fixed by choosing the normalization of the identity as $1=\lim_{a\rightarrow 0}\,V_a$. Defining then the 2-point function as the limit of the 3-point function when one of the insertions tends to the identity as in \eqref{2pnt-fn-from-3pnt-fn-space}, and using the DOZZ formula \cite{Dorn:1994xn,Zamolodchikov:1995aa} for the 3-point structure constant, the overall normalization of the 2-point function can be computed. By reflecting the DOZZ formula, 
the reflection coefficient is found to be identical to the 2-point structure constant. We hence adopt the same relation $G(\a)=\R(\a)$.
The 2-point function then reads
\be\label{2pnt-fn-TL-general R}
\langle V_{\alpha_1}\,V_{\alpha_2}\rangle=2\pi\, \frac{   \R(\alpha_1) \,\delta(\alpha_1-\alpha_2)+\delta(q-\alpha_1-\alpha_2)}{|z_{12}|^{2(\Delta_1+\Delta_2)}},
\ee
with $   G(\alpha)=  \R(\alpha)$.

We next look for the 3-point structure constant. This structure constant satisfies degenerate or shift equations \cite{Teschner:1995yf}, that result from imposing crossing symmetry of the 4-point function with insertions of the level-2 degenerate operators $V_{\langle 1,2\rangle}$, $V_{\langle 2,1 \rangle}$ \eqref{degenerate-stts}.
In particular, crossing symmetry of the 4-point function with  one insertion of the degenerate $V_{\langle 1,2\rangle}$ and three generic insertions $V_{\alpha_i}$ leads to
\begin{equation}\label{eq:AssocOPE_degen_TL}
\sum_\pm {C}(\a_1\pm\tfrac{\b}{2},\a_2,\a_3) {C}_\pm(\a_1)|  \F^s_\pm(z)|^2 =\sum_\pm {C}(\a_2\mp\tfrac{\b}{2},\a_1,\a_3) {C}_\pm(\a_2)|  \F^t_\pm(z)|^2,
\end{equation}
where $   C_{\pm}$ are the OPE coefficients
\be
V_\alpha\,V_{\langle 1,2\rangle}\sim    C_+(\alpha)\,V_{\alpha+\frac{\beta}{2}}+    C_-(\alpha)\,V_{\alpha-\frac{\beta}{2}},\qquad
V_\alpha\,V_{\langle 2,1\rangle}\sim   \tilde{C}_+(\alpha)\,V_{\alpha-\frac{1}{2\beta}}+ \tilde{  C}_-(\alpha)\,V_{\alpha+\frac{1}{2\beta}},
\ee
and $\F^s_\pm(z)$, $\F^t_\pm(z)$ are the s- and t-channel degenerate conformal blocks \eqref{conf-bloc-sch-timelike}, \eqref{conf-bloc-tch-timelike}, computed in appendix \ref{app:conformal blocks}.
Using the expressions for the degenerate fusing matrix \eqref{fusing-matrix-elements} which relates these two sets of conformal blocks, 
the above crossing symmetry equation \eqref{eq:AssocOPE_degen_TL} leads to the shift equation
\be\label{shift-eqn-3pnt-fn}
\frac{   C_+(\alpha_1)\,   C(\alpha_1+\frac{\beta}{2},\alpha_2,\alpha_3)}{   C_-(\alpha_1)\,   C(\alpha_1-\frac{\beta}{2},\alpha_2,\alpha_3)}=
\frac{\gamma\left( q\beta-2\beta\alpha_1\right)}{\gamma\left( 2\beta\alpha_1-q\beta\right)}
\prod\limits_{\pm,\pm'} \gamma\left(  \tfrac{1}{2} +\beta(\alpha_1 -\tfrac{q}{2})\pm\beta(\alpha_2 -\tfrac{q}{2})\pm'\beta(\alpha_3 -\tfrac{q}{2}) \right).
\ee
The notation $\pm,\pm'$ indicates that the two $\pm$ are independent of each other.

A second equation follows from crossing symmetry of the 4-point function with now the degenerate $V_{\langle 2,1\rangle}$ inserted, and can also be obtained from the above equation \eqref{shift-eqn-3pnt-fn} after substituting $\beta$ by $-1/\beta$ and $   C_{\pm}$ by $  \tilde{ C}_{\pm}$.

Next, we derive two analogous shift equations for the 2-point function. Crossing symmetry of the 4-point function with now two insertions of $V_{\langle 1,2\rangle}$ and two insertions of a generic operator $V_\alpha$ leads to 
\be\label{shift-eqn-2pnt-fn}
\frac{   C_+^2(\alpha)\,   G(\alpha+\beta/2)}{   C_-^2(\alpha)\,   G(\alpha-\beta/2)}=\frac{\gamma\left(q\beta -2\beta\alpha \right)\,\gamma\left( 2\beta\alpha -q \beta+\beta^2\right)}{\gamma\left(-q\beta+2\beta\alpha \right)\,\gamma\left(  -2\beta\alpha+q\beta+\beta^2\right)}.
\ee
A second shift equation again follows by using insertions of $V_{\langle 2,1\rangle}$ instead, and can be obtained from the first equation \eqref{shift-eqn-2pnt-fn} by substituting $\beta$ by $-1/\beta$ and $   C_{\pm}$ by $  \tilde{C}_{\pm}$.

The above shift equations require fixing the expression for the ratio of OPE coefficients, which effectively is part of the normalization choice. It is convenient then to separate this choice from the part of the structure constants that must satisfy the above shift relations regardless of the normalization chosen.
To this aim, we  define an operator-normalization invariant structure constant as
\be\label{definition-normalization}
   C_{\scriptscriptstyle{N}}(\alpha_1,\alpha_2,\alpha_3)\equiv \frac{   C(\alpha_1,\alpha_2,\alpha_3)}{N(\alpha_1 )\,N(\alpha_2)\,N(\alpha_3)},
\ee
where $N(\alpha_i)$ are the normalization factors coming from each operator insertion.\footnote{The primary operators $V_{\a}(z)$ are composite operators and hence need to be renormalised. The function $N(\a)$ encodes this renormalization factor, such that $N(\a)\rightarrow \lambda(\a)\,N(\a)$ under operator renormalization $V_\a\rightarrow \lambda(\a)\,V_\a$. The 3-point function $C_N(\a_i)$ is invariant under such renormalistion.}

This can now be substituted into \eqref{shift-eqn-3pnt-fn} to obtain
\begin{align}\label{shift-eqns-normalis-split}
\frac{   C_{\scriptscriptstyle{N}}(\alpha_1+\frac{\beta}{2},\alpha_2,\alpha_3)}{   C_{\scriptscriptstyle{N}}(\alpha_1-\frac{\beta}{2},\alpha_2,\alpha_3)}&=\b^{4\b(q-2\a_1)}\,\,
\prod\limits_{\pm,\pm'} \gamma\left(  \frac{1}{2} +\beta\left(\alpha_1 -\tfrac{q}{2}\right)\pm\beta\left(\alpha_2 -\tfrac{q}{2}\right)\pm'\beta\left(\alpha_3 -\tfrac{q}{2}\right) \right),\nn\\[7pt]
\frac{N^2(\alpha+\beta/2)\,\,  G(\alpha-\beta/2)}{N^2(\alpha-\beta/2)\,\,  G(\alpha+\beta/2)}&=\b^{8\b(2\a-q)}\,\,
\frac{\gamma\left( q\beta-2\beta\alpha\right)\,\gamma\left( q\beta-2\beta\alpha+\beta^2\right)}{\gamma\left( -q\beta+2\beta\alpha\right)\,\gamma\left( -q\beta+2\beta\alpha+\beta^2\right)},
\end{align}
where for the second equation we have further used \eqref{shift-eqn-2pnt-fn}. 
The relative factor of $\b^{4\b(q-2\a)}$ between the two equations is only required to be such that shifts by $\beta$ and $-1/\beta$ commute, but is otherwise arbitrary, given the ambiguity in the definitions of $N(\alpha)$ and $C_{\scriptscriptstyle{N}}$. 

The solution to the shift equation for the normalised structure constant $   C_{\scriptscriptstyle{N}}$ is given by
\be
   C_{\scriptscriptstyle{N}}(\alpha_1,\alpha_2,\alpha_3)=\Upsilon_\beta(\beta-q+\alpha_t)\,\prod\limits_{i=1}^3 \Upsilon_\beta\left(\alpha_{\scriptscriptstyle{t}}-2\alpha_i+\beta\right),
\ee
up to a constant (not determined by the shift equations) that we can choose to put to 1. The solution to the second shift equation is
\be\label{N/G}
\frac{N^2(\alpha)}{   G(\alpha)}=\frac{1}{\Upsilon_\beta(\beta+q-2\alpha)\,\Upsilon_\beta(\beta-q+2\alpha)},
\ee
again up to an $\a$-independent constant.
Given that $  G(\alpha)=  \R(\alpha)$, the left-hand side can be interpreted to be $N(\alpha)\,N(q-\alpha)$ (notice that our solution for $C_N(\a_i)$ is invariant under reflection of the charges), so it is natural to choose\footnote{In particular this means that $n(\alpha)\,n(q-\alpha)$ must be independent of $\alpha$, and proportional to the constant we have omitted in  \eqref{N/G}.}
\be
N(\alpha)=\frac{n(\alpha)}{\Upsilon_\beta(\beta+q-2\alpha)}=\frac{n(\alpha)}{\Upsilon_\beta(\beta+2\alpha)}.
\ee
It then further follows that
\be\label{2-pnt-fn-normalization-fn}
   G(\alpha)=n^2(\alpha)\,\frac{\Upsilon_\beta(\beta-q+2\alpha)}{\Upsilon_\beta(\beta+q-2\alpha)}
=n^2(\alpha)\,\beta^{2(1+\beta^2)\frac{q-2\alpha}{\beta}}\,
\frac{\Gamma\left(\beta(2\alpha-q)\right)\,\Gamma\left(\beta^{-1}(q-2\alpha)\right)}{\Gamma\left(\beta(q-2\alpha)\right)\,\Gamma\left(\beta^{-1}(2\alpha-q)\right)},
\ee
where in the second step we have used the shift relations of the Upsilon function. The function $n(\a)$ encapsulates our choice of normalization. Notice that even though it depends on $\a$, this factor in the correlators is not responsible for their behaviour under shifts, since $n(\a)$ drops out from the shift equations \eqref{shift-eqns-normalis-split}.

With the above expressions for $C_{\scriptscriptstyle{N}}(\a_i)$ and $N(\a)$, we rewrite the structure constant as
\be
   C(\alpha_1,\alpha_2,\alpha_3)=\Upsilon_\beta(\beta-q+\alpha_{\scriptscriptstyle{t}})
\prod\limits_{i=1}^3 n(\alpha_i)\,
\frac{\Upsilon_\beta(\alpha_{\scriptscriptstyle{t}}-2\alpha_i+\beta)}{\Upsilon_\beta(\beta+2\alpha_i)}.
\ee
We must now proceed to fix the normalization factor $\prod_{i=1}^3 n(\alpha_i)$. From a path integral perspective, it is clear that $n(\alpha)$ must depend on the cosmological constant $\mu$ appearing in the Liouville action.  This dependence can in fact be derived from a scaling argument in the path integral on the sphere. Indeed, upon a constant shift of the field such as $\chi\rightarrow \chi-1/2\beta\,\log(\mu)$, a correlator transforms as
\be
\langle \prod\limits_{i=1}^n V_{\alpha_i}(z_i)\rangle_\mu=\int \mathcal D \chi\,e^{-S_{tL}[\chi]}\,\prod\limits_{i=1}^n e^{-2\alpha_i\,\chi(z_i)}\rightarrow\,\mu^{\frac{\alpha_t-q}{\beta}}\,\langle \prod\limits_{i=1}^n V_{\alpha_i}(z_i)\rangle_{\mu=1}.
\ee
It is clear then that $ n(\alpha)\sim\mu^{\frac{3\alpha-q}{3\beta}}$.
The rest of the normalization can then be chosen at will.

\subsection*{Normalization}
 
One way to fix the normalization is by means of the perturbative or Coulomb gas method, explained in appendix \ref{app:DF-integrals}. We there showed that correlators whose charges satisfy $\a_t=q+n\b$ for some non-negative integer $n$, can be related to correlators in the free theory with insertions of integrated screening operators, as given by \eqref{Liouville-correls-Cg-correls-residue}. As explained in \ref{app:DF-integrals}, with such method we fix the ratio of the OPE coefficients $C_-(\a)/C_+(\a)$, and we may further fix their individual normalization since the perturbative computation shows that $C_+(\a)$ is independent of $\a$. Concretely, we found that the ratio of coefficients is
\be
\frac{C_-(\a)}{C_+(\a)}=-\frac{\pi\mu}{\gamma(\b^2)}\frac{\gamma(2\a\b+\b^2-1)}{\gamma(2\a\b)}.
\ee
This ratio can now be introduced into the shift equations for the 2- or the 3-point structure constants, which then determine the latter completely,  only up to an $\alpha_i$-independent constants. In particular, 
from equations \eqref{shift-eqn-2pnt-fn} and \eqref{2-pnt-fn-normalization-fn}, it follows that
\be
n(\a)\propto \left(\pi\mu\,\gamma(-\beta^2)\,\beta^{2+2\beta^2}\right)^{\frac{\alpha}{\beta}}.
\ee
It is clear that fixing the ratio $C_-(\a)/C_+(\a)$ is equivalent to choosing the normalization function $n(\a)$, up to an $\a$-independent constant.
In particular, this further  determines the $\alpha$-dependence of $N(\alpha)$ and of $G(\alpha)$.

Finally, we are left with determining the $\alpha_i$-independent normalization. As explained in appendix \ref{app:DF-integrals}, our choice of normalization is such that\footnote{In spacelike Liouville, the condition on the $a_i$-independent normalization of the 3-point structure constant is given by a residue condition $\underset{a_t=Q}{\text{Res}}\,\, C(a_1,a_2,a_3)= 1$, since the DOZZ formula diverges for such combinations of the charges. In the timelike case, the structure constant \eqref{TL-structure-cntt} is instead regular at $\alpha_t=q$, so we fix its value instead of the residue.}
\be
C(\alpha_1,\alpha_2,\alpha_3)\underset{\alpha_t=q}{=} -\frac{1}{2\beta}.
\ee
This condition adds a prefactor of 
\be
-\frac{1}{2\beta\Upsilon_\beta(\beta)}\left(\pi\mu\,\gamma(-\beta^2)\,\beta^{2+2\beta^2}\right)^{-\frac{q}{\beta}}
\ee
 to the 3-point structure constant. The final expression reads
\begin{align}
 C(\alpha_1,\alpha_2,\alpha_3)=-\frac{1}{2\beta}\left(\pi\mu\,\gamma(-\beta^2)\,\beta^{2+2\beta^2}\right)^{\frac{\alpha_{\scriptscriptstyle{t}}-q}{\beta}}\,
\frac{\Upsilon_\beta(\beta-q+\alpha_{\scriptscriptstyle{t}})}{\Upsilon_\beta(\beta)}
\prod\limits_{i=1}^3
\frac{\Upsilon_\beta(\alpha_{\scriptscriptstyle{t}}-2\alpha_i+\beta)}{\Upsilon_\beta(\beta+2\alpha_i)}.\nn
\end{align}

In \cite{Harlow:2011ny}, the overall normalization is given by a factor of $2\pi/\beta$ instead of $-1/2\b$. Their normalization is chosen such that the resulting 3-point function can be interpreted as arising from the standard Liouville theory path integral on an integration cycle different from that in the spacelike regime.\footnote{The factor of $2\pi/\beta$ cannot be determined by comparison with a saddle point approximation of the path integral because such terms are subleading in the semiclassical expansion. This factor is instead determined  by demanding the ratio of the DOZZ and the timelike structure constants to take a specific form consistent with the interpretation of them being computed in different integration cycles, for a generic complex value of $\b$ \cite{Harlow:2011ny}.} Our normalization instead is fixed so that it agrees with the perturbative calculations.

Finally, having fixed the normalization of the 3-point structure constant, $ \R(\alpha)$ and $ G(\alpha)=\R(\alpha)$ are fully determined, and given by \eqref{reflection-coef-timelike}.

\bibliographystyle{utphys}
\bibliography{TL}

\end{document}